\documentclass[aps,superscriptaddress,nofootinbib,floatfix,epfs,preprintnumbers]{revtex4}
\usepackage{amsfonts,amscd,amsmath,amssymb,graphicx,color,float}
\usepackage[section]{placeins}
\def\ep{\text{e}}
\def\g{\mathsf{g}}
\def\oh{\frac{1}{2}}
\def\s{\mathsf{s}}

\def\k{\mathsf{k}}
\def\n{\mathsf{n}}
\def\km{-\frac{1}{4}\ep^{\frac{1}{4}}}

\def\rq{r_q}
\def\rqq{r_{2q}}
\def\rqb{r_{\bar q}}
\def\rv{r_v}
\def\r0{r_0}
\def\rvb{r_{\bar v}}
\def\rqb{r_{\bar q}}
\def\vv{v_{\star}}

\def\QQqqb{\text{\tiny QQ}\bar{\text{\tiny q}}\bar{\text{\tiny q}}}
\def\2Qq{\text{\tiny Q}\bar{\text{\tiny Q}}{\text{\tiny q}}\bar{\text{\tiny q}}}

\def\Qqb{\text{\tiny Q}\bar{\text{\tiny q}}}

\def\QQb{\text{\tiny Q}\bar{\text{\tiny Q}}}
\def\Qqq{\text{\tiny Qqq}}

\def\QQq{\text{\tiny QQq}}
\def\QQ{\text{\tiny QQ}}
\def\Qq{\text{\tiny Qq}}
\def\3Q{3\text{\tiny Q}}

\def\qqb{\text{\tiny q}\bar{\text{\tiny q}}}

\begin{document}
\preprint{LMU-ASC 11/22}
\title{Towards a Stringy Description for  the $Q\bar Q q\bar q$-Quark System }
\author{Oleg Andreev}
 \affiliation{L.D. Landau Institute for Theoretical Physics, Kosygina 2, 119334 Moscow, Russia}
\affiliation{Arnold Sommerfeld Center for Theoretical Physics, LMU-M\"unchen, Theresienstrasse 37, 80333 M\"unchen, Germany}
\begin{abstract} 
For the case of two light flavors we propose the stringy description of the system made of one heavy and one light quark-antiquark pair, with the aim of exploring the two lower-lying Born-Oppenheimer potentials as a function of a separation of the heavy quark-antiquark pair. Our analysis reveals three critical separations related to the processes of string reconnection, breaking and junction annihilation. In particular, for the ground state potential only the process of string reconnection matters. We find that a tetraquark state makes the dominant contribution to the potential of the first excited state at small separations, and this is the big difference with the $QQ\bar q\bar q$-quark system where it does so to the ground state potential. Another big difference is the emergence of the full diquark picture $[Qq][\bar Q\bar q]$ rather than the partial picture $QQ[\bar q\bar q]$ for the tetraquark state. On the other hand, the scales of string junction annihilation, below which the systems can be thought of mainly as the compact tetraquarks, are very close for both cases and become almost the same if the phenomenological rule $E_{\QQ}=\oh E_{\QQb}$ holds. The same is also true for the screening lengths whose values are in agreement with lattice QCD.
\end{abstract}
\maketitle
\vspace{1cm}
\section{Introduction}
\renewcommand{\theequation}{1.\arabic{equation}}
\setcounter{equation}{0}

The renaissance in hadron spectroscopy came from the discovery of a new state $X(3872)$ by the Belle Collaboration \cite{X38}. In the two decades since then, more than $50$ new hadrons have been observed at high statistical significance. Most of them are potentially exotic states such as tetraquarks, pentaquarks, hybrid mesons, and glueballs.\footnote{For more details, see the recent review article \cite{lebed} and the book \cite{book}.}

One class of those exotic states includes four-quark hadrons made of one heavy and one light quark-antiquark pair. The most known example are the $Z_b$ states \cite{belle}, whose quark content is $b\bar bq\bar q$ with $q\in\{u,d\}$. Because of the large ratio of the quark masses, one of the frameworks for dealing with such a class is the Born-Oppenheimer (B-O) approximation borrowed from atomic and molecular physics \cite{bo}.\footnote{For the further development of these ideas in the context of QCD, see \cite{braat} and references therein.} In that case the corresponding B-O potentials are defined as the energies of stationary configurations of the gluon and light-quark fields in the presence of the static $Q$ and $\bar Q$ sources. The spectrum is then calculated by solving the Schr\"odinger equation in these potentials. 

Lattice gauge theory is a well-established non-perturbative approach to solving QCD. It has a long history of studying four-quark systems \cite{QQQQ} and, in particular, the B-O potentials in the $Q\bar Qq\bar q$-quark system \cite{AP, SP0}. But it still has a significant disadvantage in regards to the last problem. The available data are limited that makes it difficult to understand the physics behind them. On the other hand, the gauge/string duality \cite{uaw} provides new theoretical tools for studying strongly coupled gauge theories, and therefore may be used as an alternative method to gain important physical insights into this problem. Within this framework, the string configurations for tetraquarks were qualitatively discussed in \cite{a-3q0, coba}. Making it more precise for the case of the $Q\bar Qq\bar q$-quark system will be one of our goals. 

This paper continues our study on the doubly heavy quark systems started in \cite{a-QQq} and \cite{QQqq}. The rest of the paper is organized as follows. We begin in Sec.II by  recalling some preliminary results and setting the framework. Then in Sec.III, we construct and analyze a set of string configurations in five dimensions which provide a dual description of the $Q\bar Q q\bar q$ system. Among those we find the configurations relevant to the two lower-lying Born-Oppenheimer potentials. In addition, we comment on some other configurations and introduce three scales. These length scales characterize transitions between different dominant configurations, and in fact, are related to three types of interactions between strings: reconnection, breaking and junction (baryon vertex) annihilation. We go on in Sec.IV to compare our results with those in lattice QCD and to discuss the construction of the potentials. We conclude in Sec.V with a few important points. To make the paper more self-contained, additional technical details are included in the Appendices.


\section{Preliminaries}
\renewcommand{\theequation}{2.\arabic{equation}}
\setcounter{equation}{0}

\subsection{General procedure}

For QCD with light quarks, the B-O potentials can be defined along the line of lattice QCD. This implies that a mixing analysis based on a correlation matrix is needed \cite{FK}. Its diagonal elements are given by the energies of stationary configurations, whereas the off-diagonal ones describe transitions between those configurations. The potentials correspond to the eigenvalues of the correlation matrix. 

\begin{figure}[htbp]
\centering
\includegraphics[width=10cm]{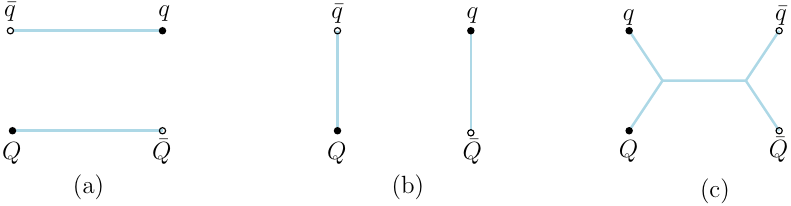}
\hspace{1.2cm}
\includegraphics[width=6.5cm]{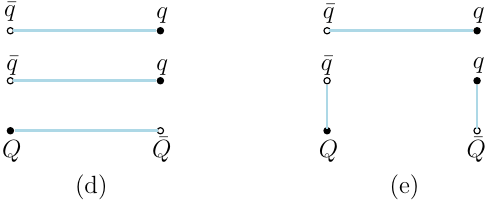}
\caption{{\small Various string configurations for the $Q\bar Qq\bar q$ system.}}
\label{c4}
\end{figure}

Now consider the string configurations for the $Q\bar Qq\bar q$-quark system from the standard viewpoint in four dimensions \cite{XA}. We do so for $N_f=2$, two dynamical flavors of equal mass, but the extension to $N_f=2+1$ is straightforward. For a first orientation, consider the simplest configurations. These are the mesonic configurations shown in Figures \ref{c4}(a) and \ref{c4}(b). Each consists of the valence quarks and antiquarks joined by the strings and looks 
like a pair of mesons. Here all the strings are in the ground state and the light quark-antiquark pair has zero momentum. Further, we can assume that other configurations are constructed by adding extra string junctions and virtual (light) quark-antiquark pairs.\footnote{Though the notion of a string junction is as old as string theory itself \cite{RV}, it took several decades to find evidence for it within lattice QCD \cite{VL}.} Intuitively, it is clear that such a procedure will result in configurations of higher energy. And so to some extent, junctions and pairs can be thought of as kinds of elementary excitations. It turns out that for our purposes we would only need relatively simple configurations. The first one is obtained by adding two junctions that leads to the tetraquark configuration of Figure \ref{c4}(c). The other two are obtained by adding a virtual pair. This is a simple modification of the meson configurations which gives rise to configurations (d) and (e), each having the three mesons. 

\begin{figure}[htbp]
\centering
\includegraphics[width=11.75cm]{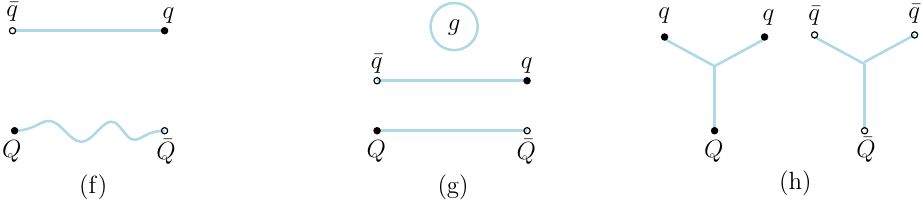}
\caption{{\small Some other configurations. The wavy line denotes an excited string and the closed string (circle) a glueball.}}
\label{c42}
\end{figure}

Before going on, it is worth making a few comments. First, the spatial positions of quarks are not significant for what follows. The only thing which matters is a separation between the heavy quark-antiquark pair. Second, apart from string junctions and virtual pairs other elementary excitations may be involved, which however are not relevant here. For example, these are some gluonic excitations: excited strings and glueball states, as sketched in Figure \ref{c42}.  
Finally, adding both elementary excitations together we can get a baryon configuration as that of Figure \ref{c42}(h). 

\begin{figure}[htbp]
\centering
\includegraphics[width=10.25cm]{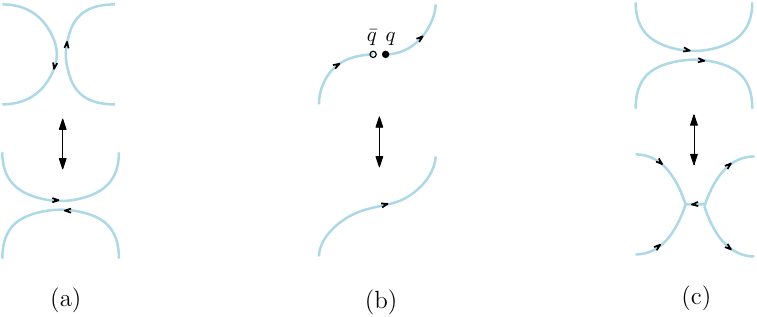}
\caption{{\small Some string interactions.}}
\label{sint}
\end{figure}

The possible transitions between the configurations arise from three type of string interactions: (a) reconnection, (b) breaking, and (c) junction annihilation (creation), as sketched in Figure \ref{sint}. 
All those are part of the big picture of QCD strings \cite{XA}. For each interaction, one can introduce a notion of a critical separation (between the heavy quark-antiquark pair) which characterizes the system. 

\subsection{Key features of a five-dimensional framework}

In our study of the $Q\bar Qq\bar q$ system, we will use the formalism recently developed in a series of papers \cite{a-strb2,a-QQq,QQqq}. Although we illustrate it using one of the simplest AdS/QCD models, the formalism is applicable to other models as well.

We consider a five-dimensional space with metric tensor 

\begin{equation}\label{metric}
ds^2=\ep^{\s r^2}\frac{R^2}{r^2}\Bigl(dt^2+d\vec x^2+dr^2\Bigr)
\,.
\end{equation}
Such a space represents a deformation of the Euclidean $\text{AdS}_5$ space of radius $R$, with a deformation parameter $\s$. There is a boundary at $r=0$ and a soft wall at $r=1/\sqrt{\s}$. Two features make it especially attractive: computational simplicity and phenomenological applications. Among those, let us only mention the model of \cite{az1} for the heavy quark potential which yields a very good fit to the lattice data \cite{white}.\footnote{For more examples, see \cite{a-hyb,a-3qPRD}.} 

To construct the configurations of Figures \ref{c4} and \ref{c42} in five dimensions, we need the building blocks. The first is a Nambu-Goto string governed by the action 

\begin{equation}\label{NG}
S_{\text{\tiny NG}}=\frac{1}{2\pi\alpha'}\int d^2\xi\,\sqrt{\gamma^{(2)}}
\,.
\end{equation}
Here $\gamma$ is an induced metric, $\alpha'$ is a string parameter, and $\xi^i$ are world-sheet coordinates. The second is a pair of string junctions, nowadays called the baryon vertices, at which three strings meet. In the AdS/CFT correspondence the baryon vertex is supposed to be a dynamic object living in ten dimensions. It is a five brane wrapped on an internal space $\mathbf{X}$ \cite{witten}, and correspondingly the antibaryon vertex is an antibrane. Both objects look point-like in five dimensions. In \cite{a-3qPRD} it was observed that the action for the baryon vertex, written in the static gauge, 

\begin{equation}\label{baryon-v}
S_{\text{v}}=\tau_v\int dt \,\frac{\ep^{-2\s r^2}}{r}
\,
\end{equation}
yields very satisfactory results, when compared to the lattice calculations of the three-quark potential. Notice that $S_{\text{v}}$ is given by the worldvolume of the brane if $\tau_v={\cal T}_5R\,\text{vol}(\mathbf{X})$, with ${\cal T}_5$ a brane tension. Unlike AdS/CFT, we treat $\tau_v$ as a free parameter to somehow account for $\alpha'$-corrections as well as possible impact of the other background fields.\footnote{Like in AdS/CFT, one expects an analog of the Ramond-Ramond fields living on the internal space $\mathbf{X}$.} In the case of zero baryon chemical potential, it is natural to take the action \eqref{baryon-v} for the antibaryon vertex so that $S_{\bar{\text{v}}}=S_{\text{v}}$.

In addition to the background metric \eqref{metric}, we introduce a background scalar field $\text{T}(r)$ which describes light quarks \cite{son}. In the present context those are at string endpoints in the interior of five-dimensional space. For our purposes it will suffice to restrict ourselves to a single scalar field (tachyon), as we are interested in the case of two light flavors of equal mass.\footnote{The use of the term tachyon seems particularly appropriate in virtue of instability of a QCD string and the worldsheet coupling to the tachyon.} The scalar field couples to the worldsheet boundary as an open string tachyon $S_{\text{q}}=\int d\tau e\,\text{T}$, where $\tau$ is a coordinate on the boundary and $e$ is a boundary metric. In what follows, we consider only a constant field $\text{T}_0$ and worldsheets whose boundaries are lines in the $t$ direction. In that case, the action written in the static gauge is simply   

\begin{equation}\label{Sq}
S_{\text q}=\text{T}_0R\int dt \frac{\ep^{\oh\s r^2}}{r}
\,.
\end{equation}
It is nothing else but the action of a point particle of mass ${\text T}_0$ at rest. Clearly, at zero baryon chemical potential the same action also describes the light antiquarks, and hence $S_{\bar{\text q}}=S_{\text q}$. 

\section{The $Q\bar Q q\bar q$-Quark System via Gauge/String Duality}
\renewcommand{\theequation}{3.\arabic{equation}}
\setcounter{equation}{0}

Now we will begin our discussion of the $Q\bar Q q\bar q$ system in the five-dimensional framework. We approach this problem from a hadro-quarkonium point of view \cite{voloshin} and hence think of the light quarks (antiquarks) as clouds. So, it makes sense to speak about their average positions or, equivalently, the centers of the clouds. The heavy quark and antiquark are thought of as point-like objects inside the clouds. Our goal is to determine the potentials as a function of the separation between the heavy quark-antiquark pair. 

An intuitive way to see the string configurations in five dimensions is to place the configurations of Figures \ref{c4} and \ref{c42} on the boundary of five-dimensional space. A gravitational force pulls the light quarks, antiquarks and strings into the interior, whereas the heavy quark and antiquark remain at rest. We start by describing configurations (a) and (b). Then we add a pair of baryon vertices or correspondingly a pair of light quarks  to those to construct configurations (c)-(e). This is a good starting point for understanding the energies of the ground and first excited states, as we will see soon.
\subsection{The disconnected configurations (a)-(b)}

First consider configuration (a). It can be interpreted as a pair of mesons. One of which is a pion, whereas the other is a quarkonium state realized by the static quark-antiquark pair connected by a string. For infinitely separated mesons, the total energy is just the sum of their masses (rest energies). It is surprising that this also holds for finite separations if one takes the average over the pion (cloud) position \cite{AP, SP0}. In what follows, we consider the configurations with pions only in the sense of averaging over all possible pion positions. 

\begin{figure}[H]
\centering
\includegraphics[width=5.25cm]{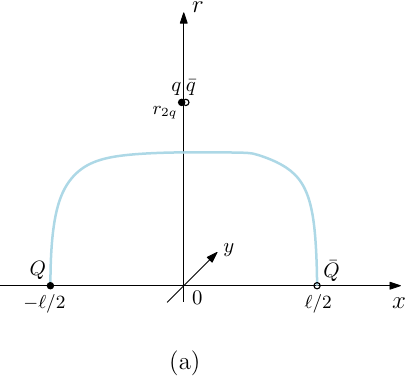}
\hspace{2.5cm}
\includegraphics[width=5.25cm]{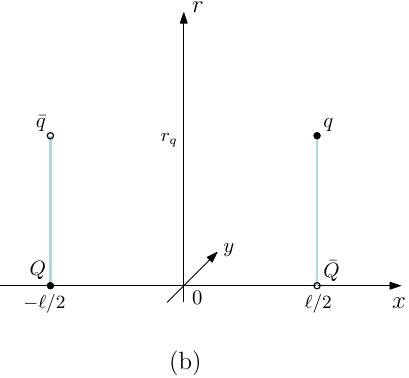}
\caption{{\small Configurations (a) and (b) in five dimensions. The heavy quark and antiquark are separated by distance $\ell$.}}
\label{con-ab}
\end{figure}

From the five-dimensional perspective, the configuration looks like that shown in Figure \ref{con-ab}(a). It consists of two 
\noindent parts: the string connecting the heavy quark-antiquark pair and the light quark pair. The first was discussed in \cite{az1}, where the relation between the string energy $E_{\QQb}$ and separation $\ell$ was described in parametric form.\footnote{For more details on this relation, see Eq.\eqref{EQQb} in Appendix C.} The second, which can be interpreted as a pion, is unknown in the literature. This part requires some explanation. In the static limit the only possible configuration is that shown in the Figure. The intuitive idea is that nothing prevents a string connecting the light quark-antiquark pair from collapsing into a point. So, the action is twice the action $S_{\text q}$. Explicitly, 

\begin{equation}\label{Sqqb}
S=2\g\n T\,\frac{\ep^{\oh\s\rqq^2}}{\rqq}
\,,
\end{equation}
with $\g=\frac{R^2}{2\pi\alpha'}$, $\n=\frac{\text{T}_0R}{\g}$, and $T=\int dt$. By varying it with respect to $\rqq$, we get

\begin{equation}\label{v3}
1-\s\rqq^2=0
\,. 
\end{equation}
This equation determines the position of the light quarks in the bulk. It has a single solution which is $\rqq=1/\sqrt{\s}$. This implies that the light quark pair is located on the soft wall. The rest energy is then  

\begin{equation}\label{Eqqb}
E_{\qqb}=2\n\sqrt{\g\sigma}
\,.	
\end{equation}
Here $\sigma$ is the physical string tension defined by \eqref{EQQb-large}. Finally, the energy of configuration (a) will be 

\begin{equation}\label{Ea}
E^{\text{(a)}}=E_{\QQb}+E_{\qqb}
\,.	
\end{equation}

Now consider configuration (b). It represents a pair of non-interacting heavy-light mesons so that the total energy is just twice that of the meson 

\begin{equation}\label{Eb}
E^{\text{(b)}}=2E_{\Qqb}
\,.	
\end{equation}
In the static limit $E_{\Qqb}$ was calculated in \cite{a-strb1} with the result

\begin{equation}\label{Qqb}
E_{\Qqb}=\g\sqrt{\s}\Bigl({\cal Q}(q)+\n \frac{\ep^{\oh q}}{\sqrt{q}}\Bigr)+c
\,.
	\end{equation}
Here the function ${\cal Q}$ is defined in Appendix A, $c$ is a normalization constant, and $q=\s\rq^2$ is a solution to the equation 

\begin{equation}\label{q}
\ep^{\frac{q}{2}}+\n(q-1)=0
\,
\end{equation}
which is nothing else but the force balance equation in the $r$-direction. It is obtained by varying the action $S=S_{\text{\tiny NG}}+S_{\text q}$ with respect to $\rq$.

\subsection{The connected configuration (c)}

It is natural to expect that if the connected configuration contributes to the ground state, or at least to  the first excited state, its shape is dictated by symmetry. If so, there are the two most symmetric case: 1. The light quarks are in the middle between the heavy quarks. 2. The light quark (antiquark) sits on top of the heavy quark (antiquark).\footnote{This is the diquark picture often used to describe the four-quark systems.} In our situation, as we will see, the former takes place at small enough separations between the heavy quark-antiquark pair, whereas the latter at larger separations. 
\subsubsection{Small $\ell$}

\begin{figure}[H]
\centering
\includegraphics[width=6.35cm]{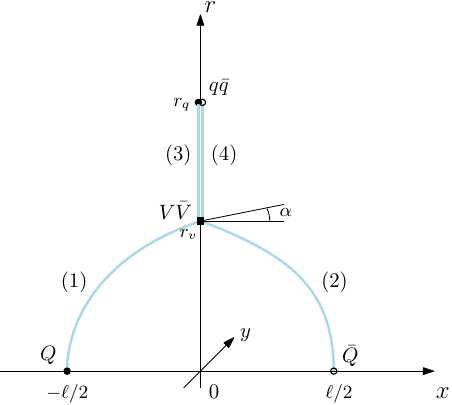}
\caption{{\small A static string configuration for small $\ell$. The light quark (antiquark) and baryon vertices are placed on the $r$-axis, respectively at $r=\rq$ and $r=\rv$. $\alpha$ indicates the tangent angle at the endpoint of the first string.}}
\label{QQqq-I}
\end{figure}

For this case, the corresponding string configuration is depicted in Figure \ref{QQqq-I}, which shows that the light quarks are indeed in the middle between the heavy ones. Here $\rq$ and $\rv$ are assumed to satisfy the following condition: $\rq>\rv$.\footnote{It indeed holds for the parameter values we are using.}  

The total action is the sum of the Nambu-Goto actions plus the actions for the vertices and light quarks

\begin{equation}\label{action-s}
S=\sum_{i=1}^4 S_{\text{\tiny NG}}^{(i)}+2S_{\text{vert}}+2S_{\text q}
\,.
\end{equation}

Picking the static gauge $\xi^1=t$ and $\xi^2=r$ for the Nambu-Goto actions, we can write the boundary conditions for $x(r)$ as follows:

\begin{equation}\label{boundary-s}
x^{(1,2)}(0)=\mp\oh\ell\,,
\qquad
x^{(i)}(\rv)=x^{(3,4)}(\rq)=0\,.
\end{equation}
The action is then\footnote{We drop the subscript $(i)$ when it does not cause confusion.}

\begin{equation}\label{action-s2}
S=2\g T
\biggl(\int_{0}^{\rv} \frac{dr}{r^2}\,\ep^{\s r^2}\sqrt{1+(\partial_r x)^2}\,\,
+
\int_{\rv}^{\rq} \frac{dr}{r^2}\,\ep^{\s r^2}\sqrt{1+(\partial_r x)^2}\,\,
+
3\k\,\frac{\ep^{-2\s\rv^2}}{\rv}
+
\n\frac{\ep^{\frac{1}{2}\s\rq^2}}{\rq}
\,\biggr)
\,,
\end{equation}
with $\k=\frac{\tau_v}{3\g}$ and $\partial_rx=\frac{\partial x}{\partial r}$.

To find a stable configuration, one has to extremize $S$ with respect to the functions $x(r)$ describing the string profiles and, in addition, with respect to $\rv$ and $\rq$ describing the locations of the baryon vertices and light quarks. As explained in Appendix B, varying with respect to $x^{(1)}$ yields 

\begin{equation}\label{l-s}
\ell=\frac{2}{\sqrt{\s}}{\cal L}^+(\alpha,v)
\,
\end{equation}
and the energy of the configuration can be written in the form 

\begin{equation}\label{E-s}
E_{\2Qq}=2\g\sqrt{\s}
\biggl(
{\cal E}^+(\alpha,v)
+
{\cal Q}(q)-{\cal Q}(v)
+
3\k\frac{\ep^{-2v}}{\sqrt{v}}
+
\n\frac{\ep^{\oh q}}{\sqrt{q}}
\biggr)
+2c
\,.
\end{equation}
Here the functions ${\cal L}^+$ and ${\cal E}^+$ are defined in Appendix A. In addition, we introduced the dimensionless variable $v=\s\rv^2$. Varying the action with respect to $\rq$ leads to Eq.\eqref{q} and with respect to $\rv$ to 

\begin{equation}\label{alpha-s}
\sin\alpha-1-3\k(1+4v)\ep^{-3v}=0
\,.
\end{equation}
Both are the force balance equations in the $r$-direction at $r=\rq$ and $r=\rv$, respectively.

Finally, the energy of the configuration is given parametrically by $E_{\2Qq}=E_{\2Qq}(v)$ and $\ell=\ell(v)$ with the parameter taking values on a subinterval of the interval $[0,1]$. Importantly, the region around $v=0$ belongs to this subinterval. This enables us to take the limit $v\rightarrow 0$ that results in ${\cal L}^+\rightarrow 0$, as follows from \eqref{fL+}. Thus, such a configuration does describe small values of $\ell$. We will return to this issue in later subsections. 

\subsubsection{Larger $\ell$}

\begin{figure}[htbp]
\centering
\includegraphics[width=6.8cm]{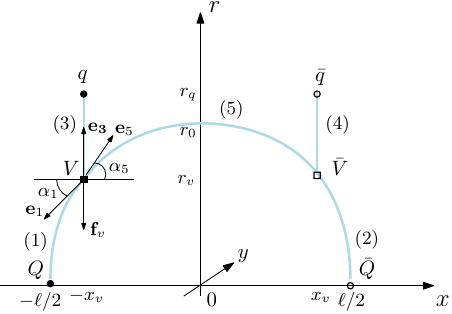}
\caption{{\small A static string configuration for larger $\ell$. The forces exerted on the baryon vertex are depicted by the arrows. $\r0$ is a turning point.}}
\label{QQqq-II}
\end{figure}

A simple numerical analysis shows that $\ell$ is an increasing function of $v$. Therefore, increasing the separation between the quark-antiquark pair will lead to a situation where the vertices reach the light quarks. In this case the configuration looks like two string meeting at a point-like defect made of the vertices and light quarks. We can continue in this way until we approach infinite separation. Why do we need to be concerned with such a configuration? The reason is that from the string theory perspective it is natural to expect that the brane-antibrane annihilation or, in other words, the string junction annihilation occurs if the positions of the vertices coincide or close enough to each other. This could lead to instability. A way out is that for these values of $\ell$ there is another configuration in which the vertices are spatially separated, as depicted in Figure \ref{QQqq-II}. It can be obtained from the configuration shown 
in Figure \ref{QQqq-I} first by splitting the baryon vertices and then by stretching a string between them.

The configuration is governed by the action 

\begin{equation}\label{action-di}
S=\sum_{i=1}^5 S_{\text{\tiny NG}}^{(i)}+2S_{\text{vert}}+2S_{\text q}
\,.
\end{equation}
In the static gauge the boundary conditions \eqref{boundary-s} are replaced by

\begin{equation}\label{bcond-d}
x^{(1,2)}(0)=\mp\oh\ell\,,
\qquad
x^{(1,2)}(\rv)=\mp x_v\,,
\qquad
x^{(3,4)}(\rv)=x^{(3,4)}(\rq)=\mp x_v\,,
\qquad
x^{(5)}(\rv)=\pm x_v
\,,
\end{equation}
so that the action now reads 

\begin{equation}\label{ac-d}
S=\g T
\biggl(
\int_{0}^{\rv} \frac{dr}{r^2}\,\ep^{\s r^2}\sqrt{1+(\partial_r x)^2}
+\int_{\rv}^{\rq} \frac{dr}{r^2}\,\ep^{\s r^2}
+\int_{\rv}^{\r0} \frac{dr}{r^2}\,\ep^{\s r^2}\sqrt{1+(\partial_r x)^2}
+ 3\k\,\frac{\ep^{-2\s\rv^2}}{\rv}
+\n\frac{\ep^{\frac{1}{2}\s\rq^2}}{\rq}
\,\biggr)
+(x\rightarrow -x)
\,.
\end{equation}
Here the integrals correspond respectively to the contributions of strings (1), (3), and (5). Because of the reflection symmetry of the configuration, only the $x<0$ part of the action is written explicitly in \eqref{ac-d}.

Varying the action with respect to $x_v$ gives\footnote{To derive it, one has to take into account the boundary conditions.} 

\begin{equation}\label{fb-x}
\cos\alpha_1-\cos\alpha_5 =0
\,. 
\end{equation}
The equation has a trivial solution $\alpha_1=\alpha_5$ which implies that strings (1) and (5) are smoothly glued together to form a single string. Another consequence of this solution is that variation with respect to $\rv$ results in 

\begin{equation}\label{v}
1+3\k(1+4v)\ep^{-3v}=0
\,.
\end{equation}
It is worth making a couple of points on this equation. First, in \cite{a-QQq} we found that in the interval $(0,1)$ it has solutions if $\k$ is restricted to the range $-\frac{\ep^3}{15} <\k\leq -\frac{1}{4}\ep^{\frac{1}{4}}$. In particular, there exists a single solution $v=\frac{1}{12}$ at $\k= -\frac{1}{4}\ep^{\frac{1}{4}}$. Second, in our analysis we have assumed that $v\leq q$. While this assumption does not hold for all possible parameter values, it holds for the ones we have in mind to use. 

Now it is straightforward to write down the energy of the configuration. It is 

\begin{equation}\label{E2Qq-d}
	E_{\2Qq}=E_{\QQb}
	+2\g\sqrt{\s}\Bigl({\cal Q}(q)-{\cal Q}(\vv)+\n\frac{\ep^{\oh q}}{\sqrt{q}}+3\k\frac{\ep^{-2\vv}}{\sqrt{\vv}}
	\Bigr)\,,
\end{equation}
with $q$ a solution to Eq.\eqref{q} and $\vv$ to Eq.\eqref{v}. In the process, we combined the contributions from strings (1), (2) and (5), to get $E_{\QQb}$ (see Appendix C). Then we performed the remaining integrals corresponding to strings (3) and (4). An important point, which applies to this expression, is that the parameter $v$ runs from $\vv$ to $1$. Thus, such a configuration does not exist for $\ell$ smaller than $\ell(\vv)$.

It is particularly useful to formally write \eqref{E2Qq-d} as follows:

\begin{equation}\label{E2Qq-di}
	E_{\2Qq}=E_{\QQb}
	+2\bigl(E_{\Qqq}-E_{\Qqb}\bigr)
	\,, 
\end{equation}
where $E_{\Qqb}$ and $E_{\Qqq}$ are defined in \eqref{Qqb} and \eqref{EQqq}. These formulas imply the occurrence of the diquark picture for the $Q\bar Q q\bar q$ system. Furthermore, it is noteworthy that the constant term in \eqref{E2Qq-di} is consistent with string breaking. 

The main conclusions of our analysis are twofold. First of all, $E_{\2Qq}$ is a piecewise function of $\ell$ or, in other words, the shape of the configuration depends on the separation between the heavy quarks. Second, the diquark picture occurs for $\ell\geq\ell(\vv)$. 

\subsubsection{Glueing the two pieces together}

Now let us discuss the gluing of the two branches of $E_{\2Qq}$. For this, we need to specify the model parameters. We will use one of the two parameter sets suggested in \cite{a-strb1} which is mainly a result of fitting the lattice QCD data to the string model we are considering. The value of $\s$ is fixed from the slope of the Regge trajectory of $\rho(n)$ mesons in the soft wall model with the geometry \eqref{metric}, and as a result, one gets $\s=0.45\,\text{GeV}^2$ \cite{a-q2}. Then, fitting the value of the string tension $\sigma$ defined in \eqref{EQQb-large} to its value in \cite{bulava} gives $\g=0.176$.\footnote{Note that this value is smaller than the value $\g=0.196$ obtained by fitting the lattice data for the heavy quark-antiquark potential in \cite{white} but the discrepancy between these two values is not significant.} The parameter $\n$ is adjusted to reproduce the lattice result for the string breaking distance in the $Q\bar Q$ system. With $\boldsymbol{\ell}_{\QQb}=1.22\,\text{fm}$ for the $u$ and $d$ quarks \cite{bulava}, this results in $\n=3.057$. 

In principle, the value of $\k$ could be adjusted to fit the lattice data for the three-quark potential, as done in \cite{a-3q} for pure $SU(3)$ gauge theory, but there are no lattice data available for QCD with two light quarks. There are still two special options: $\k=-0.102$ motivated by phenomenology\footnote{Note that $\k=-0.102$ is a solution to the equation $\alpha_{\QQ}(\k)=\oh\alpha_{\QQb}$ which follows from the phenomenological rule $E_{\QQ}(\ell)=\oh E_{\QQb}(\ell)$ in the limit $\ell\rightarrow 0$.} and $\k=-0.087$ obtained from the lattice data for pure gauge theory \cite{a-3q}. However, both are out of the range of allowed values for $\k$ as follows from the analysis of Eq.\eqref{v}. In this situation it seems natural to pick $\k=\km$ which is most close to those. 

Having fixed the parameters, we can immediately perform some simple calculations. From \eqref{q} and \eqref{v}, we get $q=0.566$ and $\vv=\frac{1}{12}$. Thus our construction of the connected configuration makes sense for these values as $q>\vv$. With this value of $v$, we find that $\ell(\vv)=0.106\,\text{fm}$. It is quite surprising that the diquark picture emerges at such small separations.  

\begin{figure}[htbp]
\centering
\includegraphics[width=8.25cm]{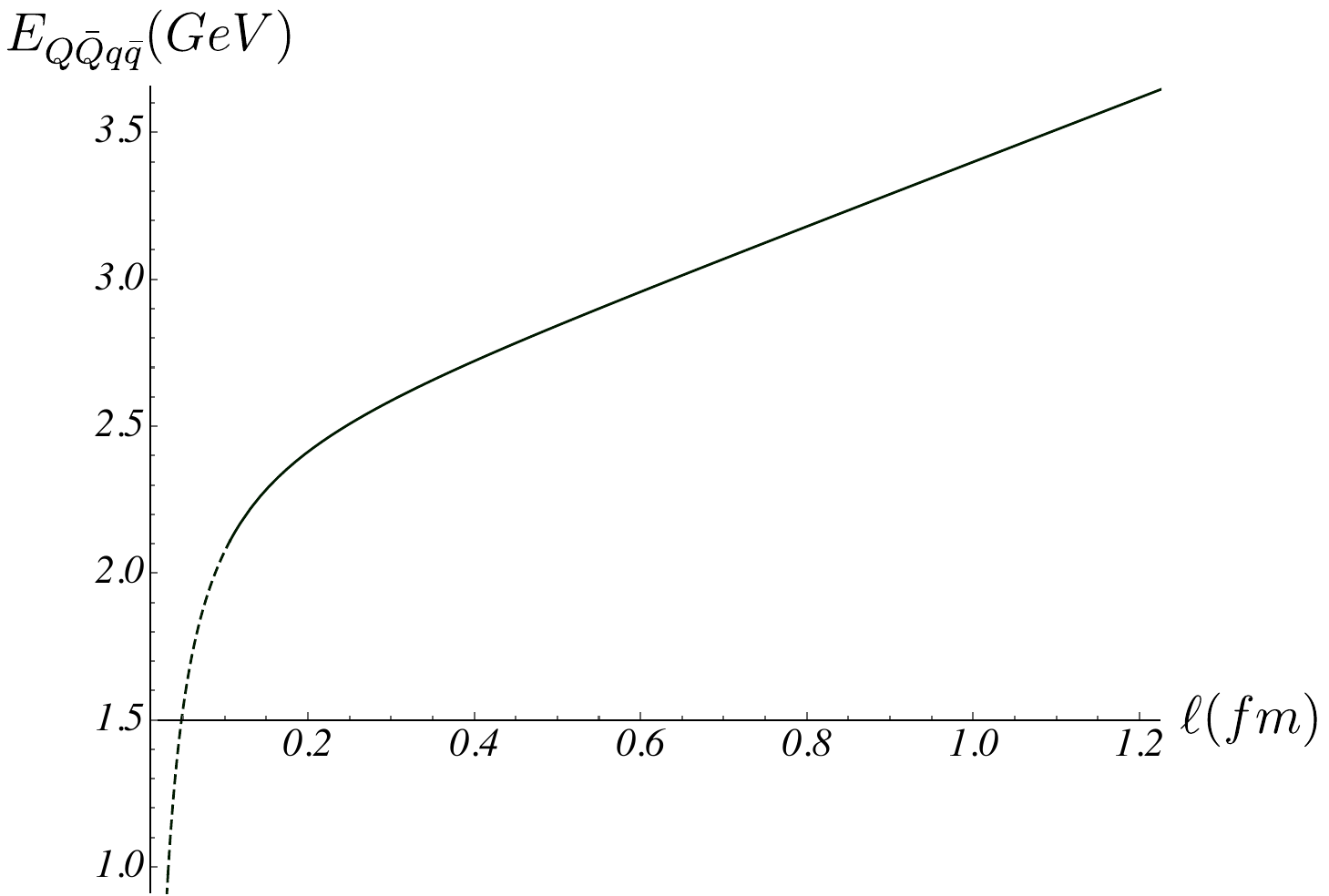}
\caption{{\small $E_{\2Qq}$ as a function of $\ell$. The dashed curve corresponds to the configuration of Figure \ref{QQqq-I} with the center pion cloud between the heavy quarks, whereas the solid curve to the configuration of Figure \ref{QQqq-II} for which the diquark picture holds. Here and later, $c=0.623\,\text{GeV}$.}}
\label{EQQqq}
\end{figure}

Given the parametric equations we derived above, it is straightforward to plot $E_{\2Qq}$ vs $\ell$. The result is presented in Figure \ref{EQQqq}. We see that the two pieces of the function $E_{\2Qq}$  are smoothly glued together at $\ell=0.106\,\text{fm}$, as expected. 
Also we will see shortly that it does behave as $1/\ell$ for small $\ell$ and linearly for large $\ell$. 
\subsubsection{The limiting cases}

We are ready to analyze the behavior of $E_{\2Qq}$ for small and large $\ell$. We begin with the case of small $\ell$. In that case, the relevant configuration is that of Figure \ref{QQqq-I}. The reason for this is that ${\cal L}^+$ (and equally $\ell$) is an increasing function of $v$ which vanishes at $v=0$. Taking the limit $v\rightarrow 0$ in Eqs.\eqref{l-s} and \eqref{E-s}, we find 

\begin{equation}\label{small}
\ell=\sqrt{\frac{v}{\s}}\Bigl(\ell_0+\ell_1v\Bigr)+o\bigl(v^{\frac{3}{2}}\bigr)
\,,\qquad
E_{\2Qq}=\g\sqrt{\frac{\s}{v}}\Bigl(E_0+E_1v\Bigr)+2E_{\Qqb}+o\bigl(v^{\frac{1}{2}}\bigr)
\,.
\end{equation}
The coefficients are given by 
\begin{gather}\label{lE}
\ell_0=\frac{1}{2}\tau^{-\frac{1}{2}}B\bigl(\tau^2;\tfrac{3}{4},\tfrac{1}{2}\bigr)
\,,
\qquad
\ell_1=\oh\tau^{-\frac{3}{2}}\Bigl(\bigl(\tau-3(1+3\k)\tfrac{\k}{\tau}\bigr)B\bigl(\tau^2;\tfrac{3}{4},-\tfrac{1}{2}\bigr)-B\bigl(\tau^2;\tfrac{5}{4},-\tfrac{1}{2}\bigr)
\Bigr)\,,
\\
E_0=2(1+3\k)+\oh\tau^{\oh}B\bigl(\tau^2;-\tfrac{1}{4},\tfrac{1}{2}\bigr)
\,,\quad
E_1=\frac{3}{2}(1+2\k)\Bigl(
-\frac{12\k}{1+3\k}
+\frac{\tau^{\oh}}{2+3\k}B\bigl(\tau^2;\tfrac{3}{4},-\tfrac{1}{2}\bigr)
-\frac{\tau^{-\oh}}{1+2\k}B\bigl(\tau^2;\tfrac{5}{4},-\tfrac{1}{2}\bigr)
\Bigr)
\,.
\end{gather}
Here $\tau=\sqrt{-3\k(3\k+2)}$ and $B(z;a,b)$ is the incomplete beta function. This makes it possible to eliminate the parameter, and thereby to get a simple result for the energy

\begin{equation}\label{E2Qq-small}
E_{\2Qq}=-\frac{\alpha_{\2Qq}}{\ell}+2E_{\Qqb}+\boldsymbol{\sigma}_{\2Qq}\,\ell+
o(\ell)
\,,	
\end{equation}
with
\begin{equation}\label{alpha-sigmaUV}
\alpha_{\2Qq}=-\g E_0\ell_0\,,
\qquad
\boldsymbol{\sigma}_{\2Qq}=\frac{1}{\ell_0}\Bigl(E_1+\frac{\ell_1}{\ell_0}E_0\Bigr)\g\s
\,.
\end{equation}
For completeness, let us estimate these coefficients. Using \eqref{alpha-QQb}-\eqref{EQQb-large}, we get $\alpha_{\2Qq}/\alpha_{\QQb}=0.939$, $\boldsymbol{\sigma}_{\2Qq}/\sigma=1.406$, with $\k$ as before. 

To find the behavior for large $\ell$, we combine Eq.\eqref{E2Qq-di} with Eq.\eqref{EQQb-large}. So,  

\begin{equation}\label{E2Qq-large}
E_{\2Qq}(\ell)=\sigma\ell+2\bigl(E_{\Qqq}-E_{\Qqb}-\g\sqrt{\s}\,I_0\bigr)+2c+o(1)
\,.
\end{equation}
This is one more indication of the universality of the string tension $\sigma$. It is the same in all the known cases: $Q\bar Q$ \cite{az1}, $QQq$ \cite{a-QQq}, $QQqq$ \cite{QQqq}, and $QQQ$ \cite{a-3q0}. 

\subsection{More disconnected configurations}

\begin{figure}[htbp]
\centering
\includegraphics[width=5.5cm]{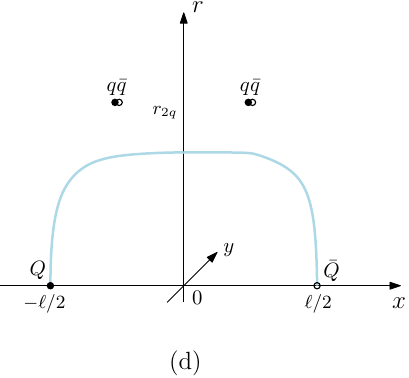}
\hspace{2cm}
\includegraphics[width=5.5cm]{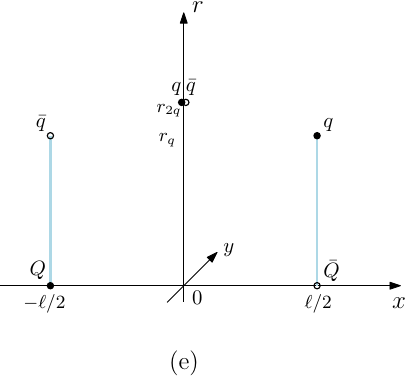}
\caption{{\small Configurations (d) and (e) in five dimensions.}}
\label{con-de}
\end{figure}

Consider configurations (d) and (e). These are obtained from configurations (a) and (b) by adding a light quark-antiquark pair (pion). In this case, the analogy with Figure \ref{con-ab} is clear: one needs to add $q\bar q$ at position $r=r_{2q}$ in the bulk. The resulting configurations are shown schematically in Figure \ref{con-de}. 
Although there are no calculations on the lattice for those, in the following we will assume that by analogy with the case of configuration (a) adding a pion, with averaging over its position, leads to an energy increase by $E_{\qqb}$. If so, then the energies can be read from the corresponding expressions for configurations (a) and (b). We have 

\begin{equation}\label{Ed}
E^{\text{(d)}}=E_{\QQb}+2E_{\qqb}
\,
\end{equation}
and 

\begin{equation}\label{Ee}
E^{\text{(e)}}=2E_{\Qqb}+E_{\qqb}
\,.	
\end{equation}
Clearly, these formulas also hold for non-interacting mesons.
\subsection{What we have learned}

\begin{figure}[htbp]
\centering
\includegraphics[width=8.8cm]{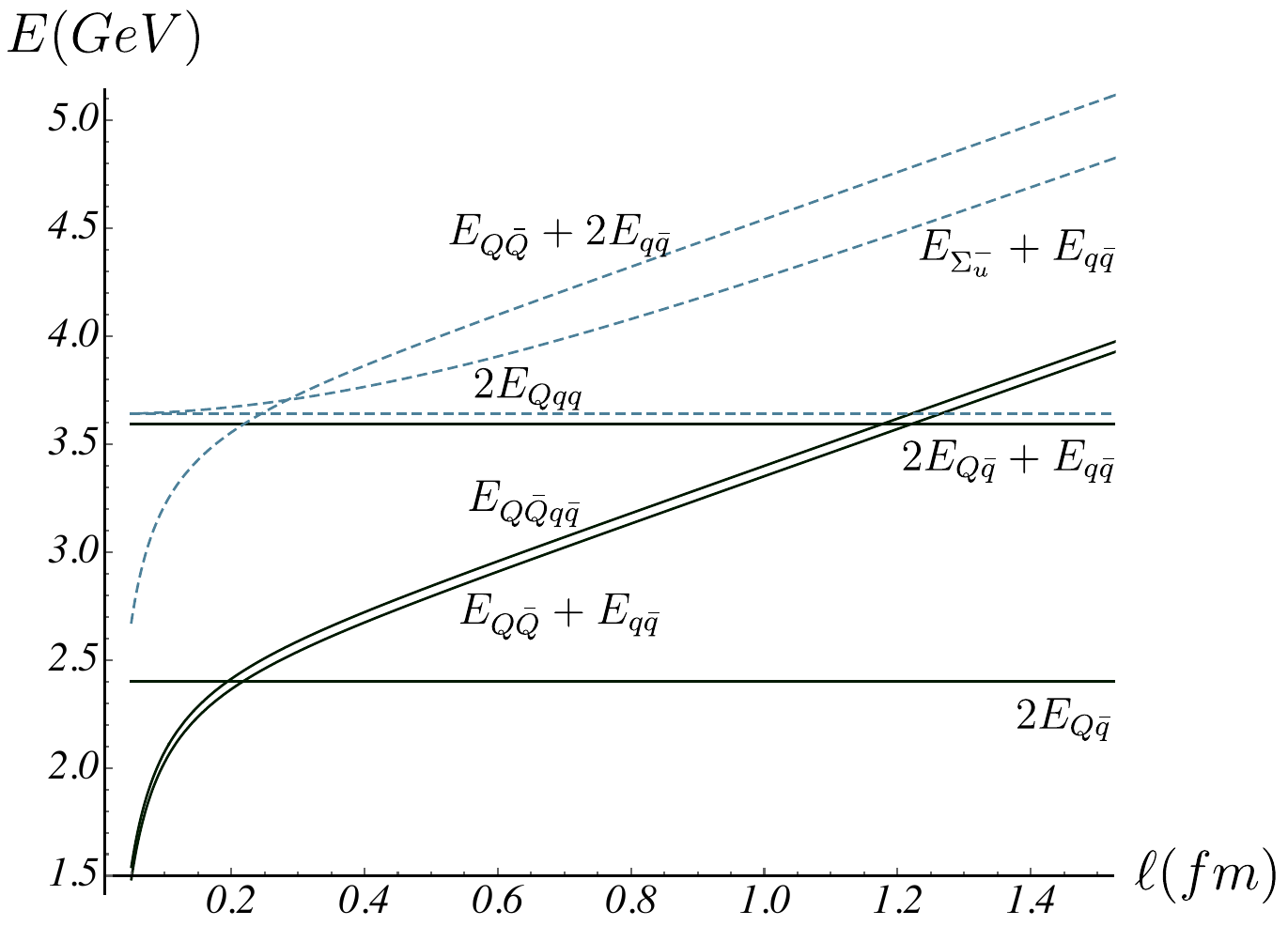}
\caption{{\small Various $E$ vs $\ell$ plots. The curves relevant for $V_0$ and $V_1$ are depicted in solid black. Here we set $\k_{\text{d}}=2000$, as in \cite{a-hyb}.}}
\label{all-L}
\end{figure}

It is instructive to see concretely how the energies of the configurations depend on the separation between the heavy quark-antiquark pair. In Figure \ref{all-L} we plot those for our parameter values. These plots show that the energies of the   

\noindent ground state and first excited state are determined only by the contributions from configurations (a)-(c) and (e): $V_0=\min\{E_{\QQb}+E_{\qqb}, 2E_{\Qqb}\}$ and $V_1=\min\{E_{\2Qq},2E_{\Qqb}, E_{\QQb}+E_{\qqb},2E_{\Qqb}+E_{\qqb}\}$.\footnote{We will discuss configurations (f) and (h) shortly.} First one has to take a minimum to get $V_0$ and then, with that in mind, $V_1$. 

An interesting consequence of the pattern of Figure \ref{all-L} is the emergence of three scales which separate different configurations, or in other words different descriptions. The first of those is a scale which refers to the process of string reconnection. It is $Q\bar Q+q\bar q\rightarrow Q\bar q+q\bar Q$ for $V_0$ and vice versa for $V_1$, if $\ell$ is increased. In the case of $V_0$ the physical meaning of this scale is that the system can be though of as a heavy quark-antiquark pair in a pion cloud for smaller quark separations and, respectively, as a pair of heavy-light meson for large ones. To make this more quantitative, we define a critical separation distance $l_{\Qq}$ by 

\begin{equation}\label{lQqb}
	E_{\QQb}(l_{\Qq})+E_{q\bar q}=2E_{\Qqb}
	\,.
	\end{equation}
If reconnection occurs at small $\ell$ (as in Figure \ref{all-L}), then by using \eqref{Qqb} and \eqref{EQQb-small}, one gets\footnote{For future reference, we keep track of $E_{\qqb}$.}

\begin{equation}\label{lQqb-small}
l_{\Qq}=
	\frac{\g\sqrt{\s}}{\boldsymbol{\sigma}_{\QQb}}
	\Bigl({\cal Q}(q)+\n\frac{\ep^{\oh q}}{\sqrt{q}}-\frac{E_{\qqb}}{2\g\sqrt{\s}}\Bigr)
	+
	\sqrt{\frac{\alpha_{\QQb}}{\boldsymbol{\sigma}_{\QQb}}+\frac{\g^2\s}{\boldsymbol{\sigma}_{\QQb}^2}
	\Bigl({\cal Q}(q)+\n\frac{\ep^{\oh q}}{\sqrt{q}}-\frac{E_{\qqb}}{2\g\sqrt{\s}}\Bigr)^2}
	\,.
\end{equation}
It is interesting to estimate $l_{\Qq}$. For the pion mass defined by \eqref{Eqqb}, the result is $l_{\Qq}=0.219\,\text{fm}$. 

The second scale is related to the process of string junction annihilation which occurs at the level of the first excited state of the system. More specifically, it is $Q\bar Qq\bar q\rightarrow Q\bar q+q\bar Q$, if $\ell$ is increased. In this case we define a critical separation distance by  

\begin{equation}\label{l2Qq}
	E_{\2Qq}(\boldsymbol{\ell}_{\2Qq})=2E_{\Qqb}
	\,.
	\end{equation}
An important observation one can make from the plots of $E_{\2Qq}$ and $2E_{\Qqb}$ is that $\boldsymbol{\ell}_{\2Qq}$ is of order $0.2\,\text{fm}$. From this, it follows that in \eqref{l2Qq} $E_{\2Qq}$ is defined by \eqref{E2Qq-di} (also see Figure \ref{EQQqq}). Since $\boldsymbol{\ell}_{\2Qq}$ is small, the equation can be solved approximately by using \eqref{EQQb-small}. So, we get 

\begin{equation}\label{L2Qq-small}
	\boldsymbol{\ell}_{\2Qq}=
	\frac{\g\sqrt{\s}}{\boldsymbol{\sigma}_{\QQb}}\Bigl({\cal Q}(\vv )-3\k\frac{\ep^{-2\vv}}{\sqrt{\vv}}\Bigr)
	+
	\sqrt{\frac{\alpha_{\QQb}}{\boldsymbol{\sigma}_{\QQb}}+\frac{\g^2\s}{\boldsymbol{\sigma}_{\QQb}^2}\Bigl({\cal Q}(\vv)-3\k\frac{\ep^{-2\vv}}{\sqrt{\vv}}\Bigr)^2}
	\,,
\end{equation}
where in the last step we used \eqref{Qqb} and \eqref{EQqq}. It is worth noting that such defined critical distance is finite and scheme independent. The normalization constant $c$ drops out of \eqref{L2Qq-small}. Moreover, it depends on $\vv$, which describes the position of the baryon vertices in the bulk, rather than $q$ which describes the position of the light quarks. This suggests that $\boldsymbol{\ell}_{\2Qq}$ is indeed related to gluonic degrees of freedom, as expected from annihilation of string junctions made of gluons. 

Let us make a simple estimate of $\boldsymbol{\ell}_{\2Qq}$. For the parameter values we use, it gives 

\begin{equation}\label{lcnum}
\boldsymbol{\ell}_{\2Qq}\approx 0.196\,\text{fm}
\,. 
\end{equation}
This values is only slightly above the value $\boldsymbol{\ell}_{\QQqqb}=0.184\,\text{fm}$ obtained for the $QQ\bar q\bar q$ system in \cite{QQqq}. Thus, our estimate provides further evidence that unlike the process of string breaking, the process of string junction annihilation takes place at relatively small scales, of order $0.2\,\text{fm}$. Of course the question arises of whether this scale is universal. We will return to this question in Sec.V, after making comparison with the lattice and gaining some information about the screening lengths.

The third scale is due to string breaking which occurs at the level of the first excited state. Here $Q\bar Q+q\bar q\rightarrow Q\bar q+q\bar Q+q\bar q$, as seen from Figure \ref{all-L}. The corresponding equation $E_{\QQb}(\ell_{\QQb})+E_{\qqb}=2E_{\Qqb}+E_{\qqb}$ reduces to Eq.\eqref{LcQQb} whose solution at large $\ell$ is given by \eqref{LQQb-large}. Numerically, it is $\ell_{\QQb}= 1.22\,\text{fm}$ \cite{bulava}. We used this value to fix the parameter $\n$ in subsection B.

\subsection{Comments on configurations (f)-(h)} 

\begin{figure}[htbp]
\centering
\includegraphics[width=5.5cm]{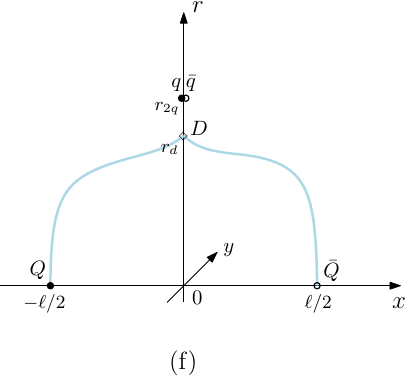}
\hspace{2cm}
\includegraphics[width=5.5cm]{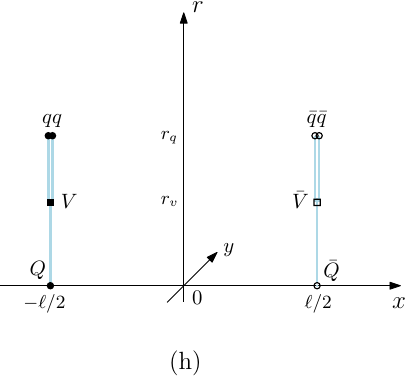}
\caption{{\small Configurations (f) and (h) in five dimensions.}}
\label{con-fg}
\end{figure}

Let us briefly discuss the role of configurations (f)-(h). The five-dimensional counterparts of configurations (f) and (h) are depicted in Figure \ref{con-fg}. In the configuration of Figure \ref{con-fg}(f) the excited string is in the $\Sigma_u^-$ state that corresponds to the $\Sigma_u^-$ hybrid potential for the 
\noindent $Q\bar Q$ system \cite{kuti}. In the current model, such a  potential was constructed from two strings meeting at a defect $D$ in the bulk \cite{a-hyb}. The configuration in Figure \ref{con-fg}(h) represents a pair of non-interacting heavy-light baryons. 

For case (f), we assume that similarly to configuration (a), its energy is the sum of two energies 

\begin{equation}\label{Ef}
E^{\text{(f)}}=E_{\Sigma_u^-}+E_{\qqb}
\,.
\end{equation}
Here $E_{\Sigma_u^-}$ is the hybrid potential and $E_{\qqb}$ is the pion mass defined as before. The potential is given in parametric form by \cite{a-hyb}

\begin{equation}\label{Ehybrid}
\ell= \frac{2}{\sqrt{\s}}{\cal L}^+(\alpha,v)
 \,,
\qquad
E_{\Sigma_u^-}=2\g\sqrt{\s}\Bigl({\cal E}^+(\alpha,v)+\k_{\text{d}}\frac{\ep^{-2v}}{\sqrt{v}}\,\Bigr)+
2c\,,
\qquad
\end{equation}
with $\sin\alpha=\k_{\text{d}} (1+4v)\ep^{-3v}$. The parameter $v$ takes values on the interval $[v_0, v_{1}]$, where $v_{0}$ is a solution to $\cos\alpha=0$ and $v_{1}$ to $\cos\alpha=v\,\ep^{1-v}$. $\k_{\text{d}}$ is a model parameter which characterizes the defect. 

It makes no sense to consider configuration (g) in detail. The basic reason that it is irrelevant for our purpose is that even for the scalar glueball, lattice calculations yield a glueball mass in the range $1.5-1.7\,\text{GeV}$ \cite{bulava2} that is larger than the physical pion mass and also $E_{\qqb}$, as we will see shortly.   

For case (h), the energy is simply

\begin{equation}\label{Eh}
E^{\text{(h)}}=2E_{\Qqq}
\,.	
\end{equation}
Here $E_{\Qqq}$ is given by \cite{a-strb1}

\begin{equation}\label{EQqq}
E_{\Qqq}=\g\sqrt{\s}\Bigl(2{\cal Q}(q)-{\cal Q}(\vv)
+2\n \frac{\ep^{\oh q}}{\sqrt{q}}
+3\k \frac{\ep^{-2\vv}}{\sqrt{\vv}}
\Bigr)+c
\,,
	\end{equation}
with $q$ a solution to \eqref{q} and $\vv$ to \eqref{v}.

Given the formulas we have just described, it is straightforward to plot the corresponding energies as a function of the separation between the heavy quark-antiquark pair. The result is presented in Figure \ref{all-L}. It is seen that these configurations contribute neither to $V_0$ nor to $V_1$, as expected. 

To complete the story, let us look at the decay channel $Q\bar Qq\bar q\rightarrow Qqq+\bar Q\bar q\bar q$. Here we define a characteristic scale by 

\begin{equation}\label{L2Qq-strb}
E_{\2Qq}(\ell)=2E_{\Qqq}
\,.	
\end{equation}
It is natural to expect that a solution to this equation is large enough. If so, then using \eqref{E2Qq-di}, we can  reduce this equation to Eq.\eqref{LcQQb} whose solution is known. It is given by $\ell_{\QQb}$ and describes the process of string breaking. In the current case, this process takes place in context of a string stretched between a diquark and an antidiquark. 


\section{More on the potentials $V_0$ and $V_1$}
\renewcommand{\theequation}{4.\arabic{equation}}
\setcounter{equation}{0}

\subsection{An issue of $E_{\qqb}$}

With the help of \eqref{Eqqb}, it is straightforward to make a simple estimate of the pion mass. Using the parameter values of Sec.III one finds that $E_{\qqb}=1.190\,\text{GeV}$. This value is considerably larger than the value $280\,\text{MeV}$ in the lattice calculations of \cite{bulava}. The problem is that the current model is not suitable for describing light hadrons as it was originally developed for hadrons with heavy quarks. What this means in practice is that at least one quark is needed to be placed on the boundary of five-dimensional space. 

\begin{figure}[htbp]
\centering
\includegraphics[width=8.35cm]{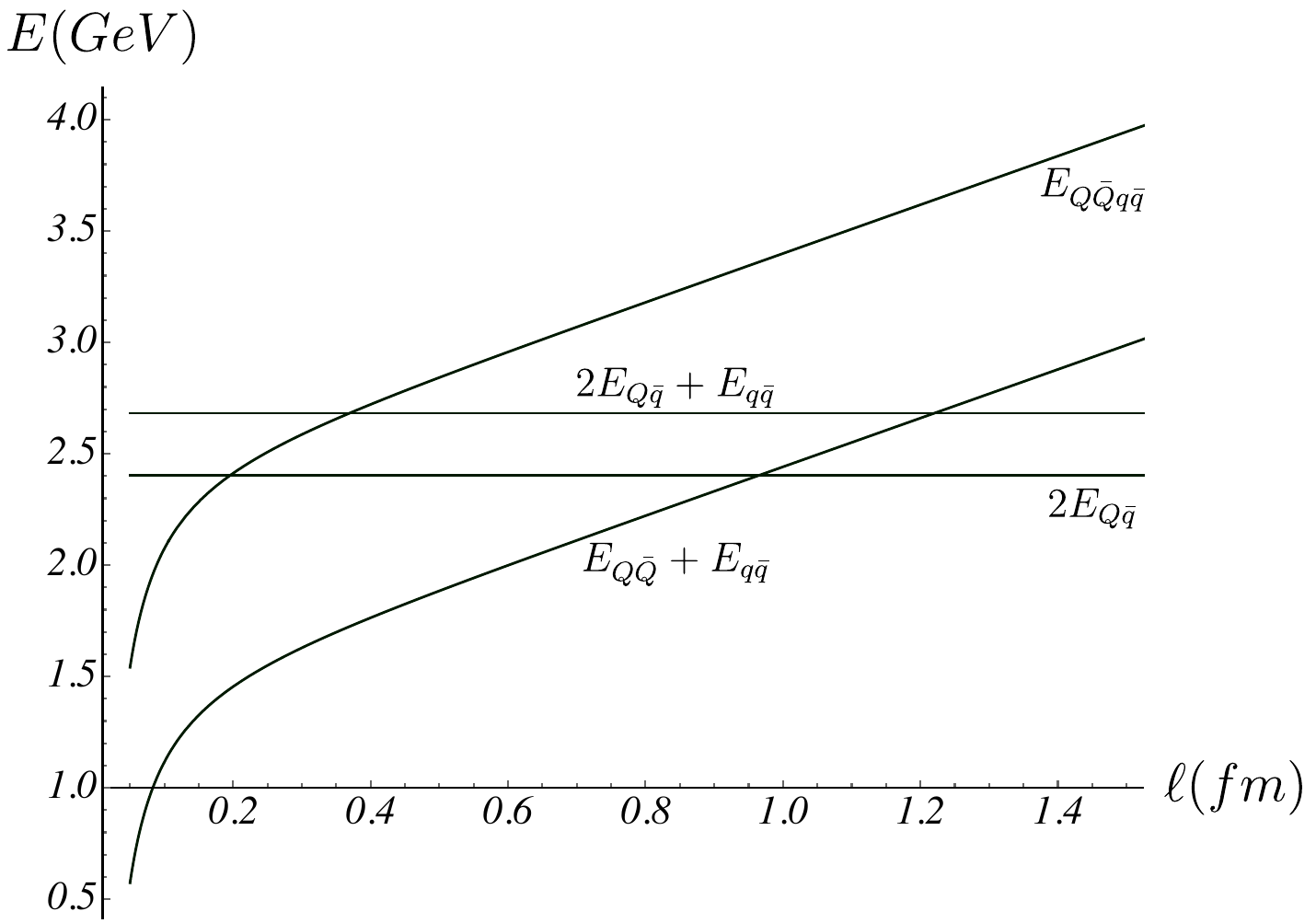}
\caption{{\small The $E$'s relevant for $V_0$ and $V_1$. Here $E_{\qqb}=280\,\text{MeV}$.}}
\label{all-pion}
\end{figure}

A possible way out is to treat $E_{\qqb}$ as a model parameter. For $E_{\qqb}=280\,\text{MeV}$, this does not cause an essential change of the plots of Figure \ref{all-L}, as illustrated in Figure \ref{all-pion}. 

The most visible effect of change in $E_{\qqb}$ is that string reconnection occurs at much larger $\ell$. Thus, in order to solve the equation \eqref{lQqb}, one now has to use the asymptotic expansion \eqref{EQQb-large}. A simple algebra yields 

\begin{equation}\label{lQq-large}
\ell_{\Qq}=
\frac{2\g\sqrt{\s}}{\sigma}
\Bigl(Q(q)+\n\frac{\ep^{\frac{q}{2}}}{\sqrt{q}}	+I_0-\frac{E_{\qqb}}{2\g\sqrt{\s}}\,\Bigr)
\,.
\end{equation}
For $E_{\qqb}=280\,\text{MeV}$, $\ell_{\Qq}=0.964\,\text{fm}$. For $E_{\qqb}=496\,\text{MeV}$, $\ell_{\Qq}=0.766\,\text{fm}$ that agrees with the lattice result \cite{AP}.  

An interesting conclusion can be drawn from the examples given: Our results indicate that the following scale hierarchy is met 

\begin{equation}\label{scales}
\boldsymbol{\ell}_{\2Qq}< l_{\Qq}< \ell_{\QQb}
\,.
\end{equation}
At this point, one may well ask what happens at the physical value of the pion mass. We expect that at least the string junction annihilation distance remains the smallest scale, namely $\boldsymbol{\ell}_{\2Qq}< l_{\Qq}$, $\ell_{\QQb}$. Of course, it is not the whole answer to this question, but it's certainly an important piece of it.

\subsection{Comparison with the lattice}

The potentials for the $Q\bar Qq\bar q$ system have been studied in lattice QCD \cite{AP,SP0}. In this case those are extracted from the set of correlators. The data limitation prevents one from being able to analyze the effects of string breaking and reconnection but not the effect of string junction annihilation. In particular, the data extracted from the correlator of meson operators allow one to gain some insight into the potential $V_1$ in the quark separation range $0.1\text{-}0.7\,\text{fm}$ \cite{AP}. They are fitted by a single algebraic expression

\begin{equation}\label{latticeV1}
V_{1}(\ell)=-\frac{\alpha}{\ell}\exp\Bigl(-\frac{\ell^{\,p}}{d^{\, p}}\,\Bigl)\,+2 E_{\Qqb}
\,\,,	
\end{equation}
with the parameters $\alpha$, $d$, and $p$. 

Keeping in mind the fact that $\boldsymbol{\ell}_{\2Qq}$ is $0.196\,\text{fm}$, we expect that the screening length $d$ is near this value. Therefore we use \eqref{E2Qq-di} for $V_1$, and then solve for the unknown coefficients in the limit of small $\ell$. So, we get 

\begin{equation}\label{dp}
	\alpha=\alpha_{\QQb}
	\,,\qquad
	d=\sqrt{\frac{\alpha_{\QQb}}{\boldsymbol{\sigma}_{\QQb}}}
	\,,\qquad
	p=2
	\,.
\end{equation}
The parameters are expressed in terms of the coefficients of the quark-antiquark potential. We can make an estimate of the screening length. With our parameter set, this gives $d=0.238\,\text{fm}$. The value is close to the range of lattice QCD, where $d=0.16^{+0.05}_{-0.02}\,\text{fm}$ \cite{AP,wagner}.

\begin{figure}[H]
\centering
\includegraphics[width=8.2cm]{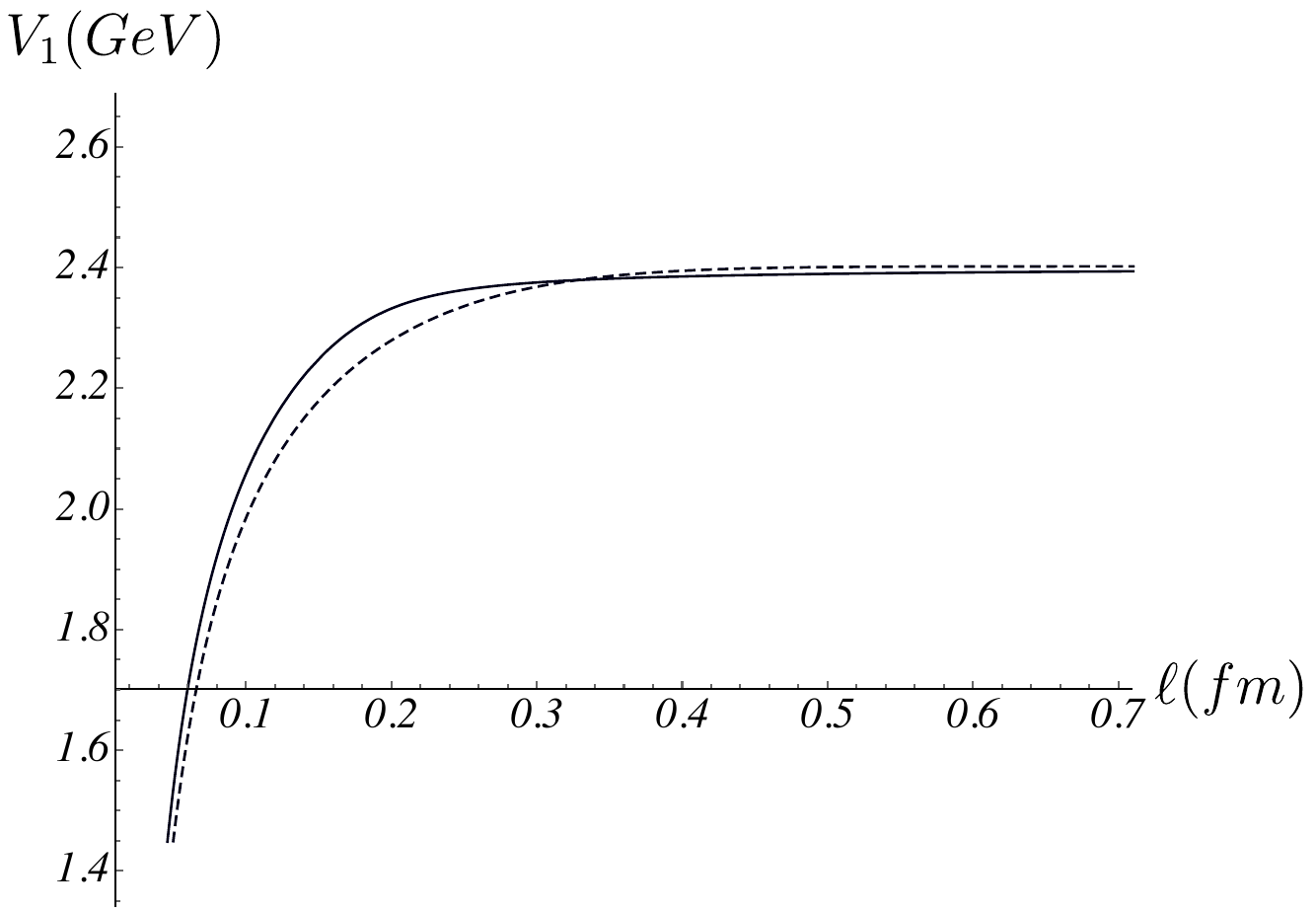}
\hspace{0.75cm}
\includegraphics[width=8.75cm]{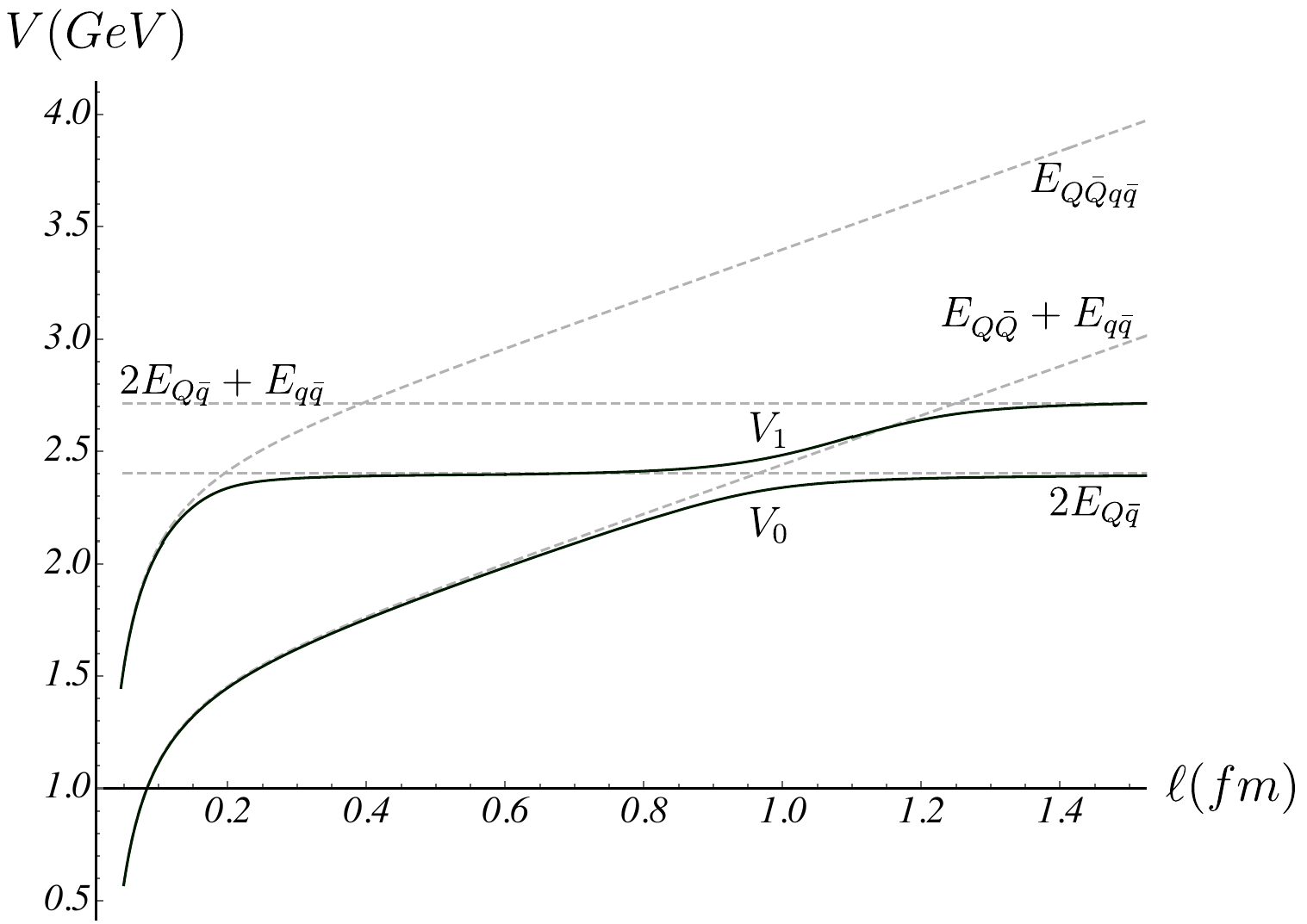}
\caption{{\small Left: The potential $V_1$ defined by the model Hamiltonian \eqref{HV1}. Here $\Theta=75\,\text{MeV}$. The dashed curve corresponds to the lattice parameterization \eqref{latticeV1}. Right: The potentials $V_0$ and $V_1$. The relevant $E$'s are shown in dashed lines. Here $E_{\qqb}=280\,\text{MeV}$.}}
\label{MW}
\end{figure}

 What makes the lattice approach less attractive is that it is difficult to say whether the system can be thought of as a compact tetraquark or as a pair of mesons. As we have just seen, this question can be answered in the framework of the string model. Now for completeness let us show how to construct the potential $V_1$ in this range of quark separations. Consider the model Hamiltonian 

\begin{equation}\label{HV1}
{\cal H}(\ell)=
\begin{pmatrix}
E_{\2Qq}(\ell) & \Theta \\
\Theta & 2E_{\Qqb} \\
\end{pmatrix}
\,,
\end{equation}
where $\Theta$ describes the strength of the mixing between the compact tetraquark state and two mesons. The potential is given by the smallest eigenvalue of ${\cal H}$. Explicitly, 

\begin{equation}\label{V1MW}
V_1=\oh\Bigl(E_{\2Qq}+2E_{\Qqb}\Bigr)
-
\sqrt{\frac{1}{4}\Bigl(E_{\2Qq}-2E_{\Qqb}\Bigr)^2+\Theta^2}
\,.	
\end{equation}
In principle, $\Theta$ can be computed from a correlator of the operators corresponding to these two states, but currently this is out of our reach. So, we treat $\Theta$ as a free parameter and find its value by the best fit of our prediction to the parameterization suggested from lattice QCD. In Figure \ref{MW} on the left we plot $V_1$ vs $\ell$. There are two things worth  
\noindent mentioning here. The first is a visible deviation between the solid and dashed curves on the interval $0.08\lesssim\ell \lesssim 0.28\,\text{fm}$. In our opinion, it is early to speak about the real reasons for this as the lattice data are very limited. The second is a falloff at large $\ell$. It is power law for \eqref{V1MW} with $\Theta=const$, but exponential for \eqref{latticeV1}. Of course, one can get the desired exponential falloff by taking $\Theta$ as a Gaussian function with a pick at $\ell=\boldsymbol{\ell}_{\2Qq}$ but it will require one addition parameter (a Gaussian width). 

\subsection{The potentials}

Having understood the string configurations relevant to the ground and first excited states of the $Q\bar Q q\bar q$ system, one can gain some insight into their energies. One way for doing so is to consider a model Hamiltonian, similar as it is used in lattice QCD to study the phenomenon of string breaking \cite{FK}. For the problem at hand it is  

\begin{equation}\label{HV0V1}
{\cal H}(\ell)=
\begin{pmatrix}
\,E_{\QQb}(\ell)+E_{\qqb} & {} & {} & {} & {}\, \\
\,{} & 2E_{\Qqb} & {} & {} & \Theta_{ij} \,\\
\,\Theta_{ij}& {} & E_{\2Qq}(\ell) & {} & {}\, \\
\,{} & {} & {} & {} & 2E_{\Qqb}+E_{\qqb}\,\\
\end{pmatrix}
\,,
\end{equation}
where the off-diagonal elements describe the strength of mixing between the four states (string configurations). The potentials of interest are the two smallest eigenvalues of the matrix ${\cal H}$. 

Unlike lattice QCD, where the Hamiltonian could be extracted from a correlation matrix, it is not clear how to compute the off-diagonal elements within the string models. This makes it difficult to see concretely how the potentials look like. Nevertheless, we can learn from our experience in the previous subsection (see also Appendix C) about the order of magnitude of the $\Theta$'s near the intersection points. With the help of this, the picture will then look more like what is shown in Figure \ref{MW} on the right. A compact tetraquark configuration contributes dominantly to the $V_1$ potential at small quark separations, as we have seen this before.

\section{Concluding comments}
\renewcommand{\theequation}{5.\arabic{equation}}
\setcounter{equation}{0}

(1) There is really a difference between the $Q\bar Q q\bar q$ and $QQ\bar q\bar q$-quark systems. In the former, a compact tetraquark structure shows up in $V_1$, while in the latter in $V_0$. The basic reason is that a pair $Q\bar Q$ being a color singlet enables configuration (a) which dominates at small heavy quark separations and hence contributes to $V_0$. Nevertheless the estimate of the corresponding scales shows that the value of $\boldsymbol{\ell}_{\2Qq}$ is small and close to that of $\boldsymbol{\ell}_{\QQqqb}\,$, see \eqref{lcnum}. Interestingly, these values become almost the same if the phenomenological rule $E_{\QQ}=\oh E_{\QQb}$ holds true.\footnote{In the literature it is sometimes called the Lipkin rule.} To see this, let us first note that the equation $E_{\2Qq}(\ell)=2E_{\Qqb}$ reduces to $\oh E_{\QQb}(\ell)=2E_{\Qqb}-E_{\Qqq}$ as follows from Eq.\eqref{E2Qq-di}. On the other hand, the heavy quark-diquark symmetry implies that the equation $E_{\QQqqb}(\ell)=2E_{\Qqb}$ for small $\ell$ becomes $E_{\QQ}(\ell)\approxeq2E_{\Qqb}-E_{\Qqq}$. Thus 

\begin{equation}\label{LipkinLc}
\boldsymbol{\ell}_{\2Qq}\approxeq\boldsymbol{\ell}_{\QQqqb}
\end{equation}
if the Lipkin rule holds. The same is true for the screening lengths. Those are given respectively by $d=\sqrt{\frac{\alpha_{\QQb}}{\boldsymbol{\sigma}_{\QQb}}}$ and $d=\sqrt{\frac{\alpha_{\QQ}}{\boldsymbol{\sigma}_{\QQ}}}$ \cite{QQqq}. The $\alpha$'s and $\boldsymbol{\sigma}$'s are the coefficients in the small $\ell$ expansions of $E_{\QQb}$ and $E_{\QQ}$ (see Eq.\eqref{EQQb-small}). Clearly, rescaling of the coefficients by the factor 2 has no effect on the $d$'s. 

(2) An interesting experimental observation is that the $Z$ states are very close to the corresponding $B\bar B$ thresholds \cite{ali}. Moreover, the mass differences with respect to the thresholds are positive.\footnote{This requires a caveat because the errors in the mass determination are not small enough.} If so, then there are no binding energies which bind the mesons in such states. This could seem puzzling for one who thinks only in terms of $V_0$, but not for one who also takes into consideration $V_1$ (see the right panel of Figure \ref{MW}). Thus, the story may really and truly be about the four states contributing to the construction of $V_1$, one of which is a compact tetraquark. 

(3) The diquark picture emerged in the case of configuration (c) gives rise to a naturally defined diquark mass. Using the $\ell$-independent part in Eq.\eqref{E2Qq-d}, we define it by\footnote{For this definition to make sense, the separation between diquarks must be larger than $\ell(\vv)$.}
 
\begin{equation}\label{Mdi}
	m_{[Qq]}=m_Q
	+\g\sqrt{\s}\Bigl({\cal Q}(q)-{\cal Q}(\vv)+\n\frac{\ep^{\oh q}}{\sqrt{q}}+3\k\frac{\ep^{-2\vv}}{\sqrt{\vv}}
	\Bigr)\,,
\end{equation}
with $q$ and $\vv$ as in \eqref{E2Qq-d}. The square brackets mean that a diquark has zero spin. It is interesting to see what results come out of this definition and how those agree or disagree with phenomenological values. There is a problem here. Though the masses of the light quarks in our model are known (see the discussion below), we can say nothing specific about the masses of the heavy quarks because of the lack of lattice data \cite{bulava}. This makes it impossible to directly estimate diquark masses via the above formula. But what we can do is to check the prediction that the mass difference between a diquark and a heavy quark is independent of the mass of the latter. To this end, we consider some phenomenological models \cite{DE,Gi,lu}. The results are presented in Table \ref{estimates}. 
\begin{table*}[htb]
\renewcommand{\arraystretch}{2}
\centering 	
\begin{tabular}{lccccccr}				
\hline
Diquark       ~~&~~~ \cite{DE}  ~~~&~~~~~\cite{Gi} ~~~&~~~~\cite{lu}~~~
\rule[-3mm]{0mm}{8mm}
\\
\hline 
$[cq]$ & 423  & ~~387 & ~471  \\
$[bq]$ & 479 & ~~374 & ~474  \\
\hline
\end{tabular}
\caption{ \small The mass difference $m_{[Qq]}-m_Q$ (in MeV).}
\label{estimates}
\end{table*}
Although there is a visible discrepancy in the case of \cite{DE}, it is about $3\%$ in \cite{Gi} and $0.6\%$ in \cite{lu}.

(4) For the parameter values used here, the masses of the light quarks can be found by fitting the string breaking distance \eqref{LQQb-large} to the lattice data of \cite{bulava}. It leads to the result $m_{u/d}=46.6\,\text{MeV}$ \cite{a-stb3q}. Thus, the value falls between the usual values of the current and constituent quark masses. This however requires some caveats. First, the lattice calculations were done at unphysical pion mass $m_{\pi}=280\,\text{MeV}$. Second, the main assumption of \cite{a-stb3q} is that ${\text T}_0=m$, with $T_0$ a constant. This is a simplified version of the assumption of \cite{son} which doesn't include the effect of the chiral condensate. Finally, it is worth mentioning that the use of the more phenomenologically motivated parameter values yields $m_{u/d}=23.5\,\text{MeV}$ \cite{a-stb3q}. 

(5) Actually, our discussion has been so general that it makes sense with or without spin. But the neglect of spin-dependent effects is not good in making contact with the real world. In particular, we ran into trouble with $q\bar q$, where one expects such effects. Also including spin will be crucial in predicting the masses of spin one diquarks. 

(6) Intuitively, one expects the probability of reconnection between the strings connecting quarks in $Q\bar Q$ and $q\bar q$ to increase with increasing a separation of the heavy quark-antiquark pair. But it is not clear at what separation the probability reaches a maximum.\footnote{It is natural to expect that such a separation coincides with the critical separation $l_{\Qq}$ defined from the energy balance equation \eqref{lQqb}.} Let us assume that this happens at the maximum possible separation of the $Q\bar Q$   
\begin{figure}[htbp]
\centering
\includegraphics[width=3.5 cm]{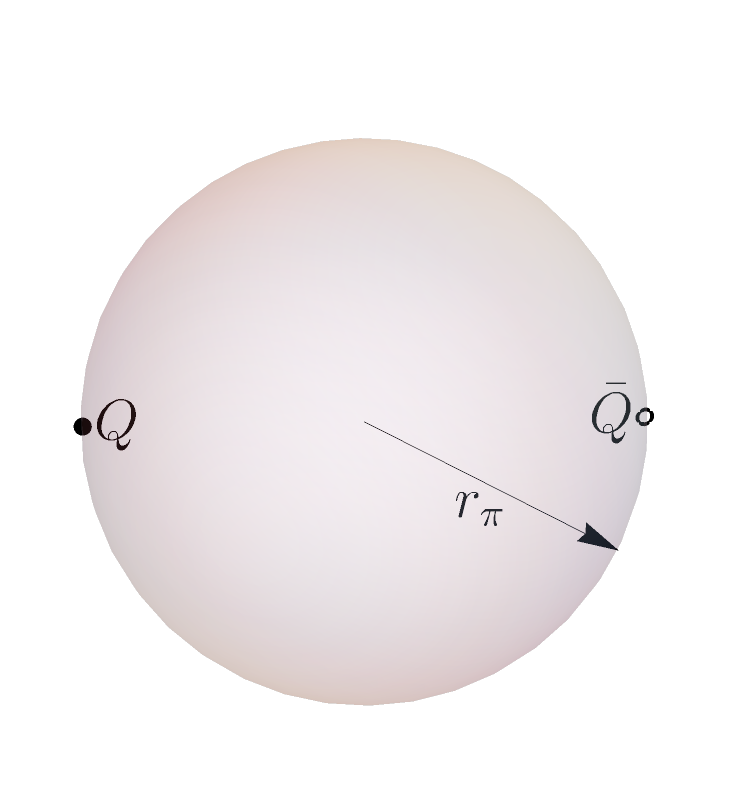}
\caption{{\small A heavy quark-antiquark pair inside a pion cloud of radius $r_{\pi}$.}}
\label{rpion}
\end{figure}

\noindent pair inside the pion cloud, as sketched in Figure \ref{rpion}. If so, then 

\begin{equation}\label{lcpion}
l_{\Qq}=2r_{\pi}
\,.
\end{equation}
For $m_{\pi}=265\,\text{MeV}$, which is quite close to the value in \cite{bulava}, the (charge) pion radius is $\langle r_\pi^2\rangle =0.232\,\text{fm}^2$ \cite{pionR}. So, we immediately obtain $l_{\Qq}=0.963\,\text{fm}$. Although there is no satisfactory explanation of why our assumption is correct, it is interesting that it results in approximately the same value of $l_{\Qq}$ as Eq.\eqref{lQq-large} evaluated at $m_{\pi}=280\,\text{MeV}$. 

Eq.\eqref{lcpion}, if taken literally, gives $l_{\Qq}=1.318\,\text{fm}$ at $\langle r_\pi^2\rangle =0.434\,\text{fm}^2$ quoted by Particle Data Group \cite{PDG}. This then implies that, in the real world, one might expect $\ell_{\QQb}<l_{\Qq}$. So the hierarchy of scales in \eqref{scales} has to be modified, but we will omit this from the discussion.

(7) The $Q\bar Qq\bar q$-quark system has a rich complexity of physics and a number of unanswered, pressing questions. Making further progress in theoretical understanding and applications to hadron phenomenology will require a joint effort by the high-energy community. We hope that our study provides a useful starting point for a stringy approach to this system.

\begin{acknowledgments}
We would like to thank S. Dubovsky, A. Peters, J.-M. Richard, J. Sonnenschein, and M. Wagner for useful communications and conversations. This work is supported by Russian Science Foundation grant 20-12-00200 in association with Steklov Mathematical Institute. We acknowledge the warm hospitality at the Simons Center for Geometry and Physics at Stony Brook University and at the Center for Cosmology and Particle Physics at New York University, where part of this work was completed.
\end{acknowledgments}
\appendix
\section{Notation and definitions}
\renewcommand{\theequation}{A.\arabic{equation}}
\setcounter{equation}{0}
In all Figures throughout the paper, heavy and light quarks (antiquarks) are denoted by $Q\,(\bar Q)$ and $q\,(\bar q)$, and baryon (antibaryon) vertices by $V\,(\bar V)$. Strings are represented by smooth curves without self-intersections. When not otherwise noted, we usually set light quarks (antiquarks) at $r=\rq\,(\rqb)$ and vertices at $r=\rv\,(\rvb)$. For convenience, we introduce dimensionless variables: $q=\s\rq^2$, $\bar q=\s\rqb^2$, $v=\s\rv^2$, and $\bar v=\s\rvb^2$. They take values on the interval $[0,1]$ and show how far from the soft-wall these objects are.\footnote{In these dimensionless units, the soft wall is located at $1$.} To classify the critical separations related to the string interactions of Figure \ref{sint}, the notation $l$ is used for (a), $\ell$ for (b), and $\boldsymbol{\ell}$ for (c).

In order to present the formulas in a general simplified form, we use the set of basic functions \cite{a-stb3q}: 

\begin{equation}\label{fL+}
{\cal L}^+(\alpha,x)=\cos\alpha\sqrt{x}\int^1_0 du\, u^2\, \ep^{x (1-u^2)}
\Bigl[1-\cos^2{}\hspace{-1mm}\alpha\, u^4\ep^{2x(1-u^2)}\Bigr]^{-\frac{1}{2}}
\,,
\qquad
0\leq\alpha\leq\frac{\pi}{2}\,,
\qquad 
0\leq x\leq 1
\,.
\end{equation}
It is a non-negative function which vanishes if $\alpha=\frac{\pi}{2}$ or $x=0$, and has a singular point at $(0,1)$;

\begin{equation}\label{fE+}
{\cal E}^+(\alpha,x)=\frac{1}{\sqrt{x}}
\int^1_0\,\frac{du}{u^2}\,\biggl(\ep^{x u^2}
\Bigl[
1-\cos^2{}\hspace{-1mm}\alpha\,u^4\ep^{2x (1-u^2)}
\Bigr]^{-\frac{1}{2}}-1-u^2\biggr)
\,,
\qquad
0\leq\alpha\leq\frac{\pi}{2}\,,
\qquad 
0\leq x\leq 1
\,.
\end{equation}
This function is singular at $x=0$ and $(0,1)$;

\begin{equation}\label{Q}
{\cal Q}(x)=\sqrt{\pi}\text{erfi}(\sqrt{x})-\frac{\ep^x}{\sqrt{x}}
\,,
\end{equation}
which is the special case of ${\cal E}^+$ obtained by setting $\alpha=\frac{\pi}{2}$. Here  $\text{erfi}(x)$ denotes the imaginary error function. A useful fact is that the small $x$ behavior is 

\begin{equation}\label{Q0}
{\cal Q}(x)=-\frac{1}{\sqrt{x}}+\sqrt{x}+O(x^{\frac{3}{2}})
\,;
\end{equation}

\begin{equation}\label{I}
	{\cal I}(x)=I_0 -
\int_{\sqrt{x}}^1\frac{du}{u^2}\ep^{u^2}\Bigl[1-u^4\ep^{2(1-u^2)}\Bigr]^{\frac{1}{2}}\,
\,,
\qquad
I_0=\int_0^1\frac{du}{u^2}\Bigl(1+u^2-\ep^{u^2}\Bigl[1-u^4\ep^{2(1-u^2)}\Bigr]^{\frac{1}{2}}\Bigr)
\,,\qquad
0< x\leq 1
\,.
\end{equation}
Numerically, $I_0\approx 0.751$.

\section{A static Nambu-Goto string with fixed endpoints}
\renewcommand{\theequation}{B.\arabic{equation}}
\setcounter{equation}{0}

The purpose of this Appendix is to briefly describe some facts about a static Nambu-Goto string in the curved geometry \eqref{metric} that may be helpful for understanding the string configurations of Sect.III. For our purposes, the only 
\begin{figure}[htbp]
\centering
\includegraphics[width=4.15cm]{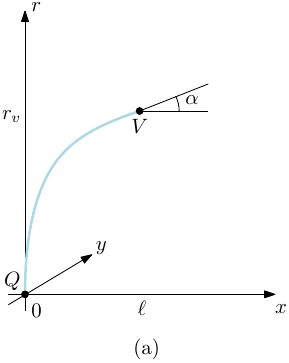}
\hspace{3.5cm}
\includegraphics[width=3.55cm]{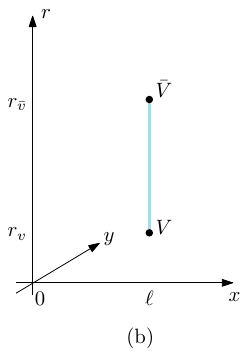}
\caption{{\small A static string stretched between two points. $\alpha$ is the tangent angle. (a) $0<\alpha<\frac{\pi}{2}$. (b) $\alpha=\frac{\pi}{2}$.}}
\label{ng}
\end{figure}
cases we need to consider are presented in Figure \ref{ng}. More details and further results omitted in this Appendix can be found in \cite{a-3qPRD}.

Consider a string stretched between the two fixed points $Q(0,0)$ and $V(\ell,\rv)$ in the $xr$-plane as shown in Figure \ref{ng}(a). To get to the desired formulas here as quickly as possible, we will take as in \cite{malda} the static gauge $\xi_1 = t$ and $\xi_2 = x$. In this case the string profile is described by the function $r(x)$ subject to the boundary conditions

\begin{equation}\label{ng-bc}
r(0)=0\,,\qquad r(\ell)=\rv
\,.
\end{equation}
The Nambu-Goto action takes the form

\begin{equation}\label{NG+}
S=T\g\int_{0}^{\ell} dx\, w(r)\sqrt{1+(\partial_x r)^2}
\,,\qquad
w(r)=\frac{\ep^{\s r^2}}{r^2}
\,.
\end{equation}
Here $\g=\frac{R^2}{2\pi\alpha'}$, $T=\int dt$, and $\partial_x r=\frac{\partial r}{\partial x}$. Since the integrand does not explicitly depend on $x$, the corresponding Euler-Lagrange equation has the first integral

\begin{equation}\label{In}
I=\frac{w(r)}{\sqrt{1+(\partial_x r)^2}}\,.
\end{equation}

After expressing $I$ in terms of $\alpha$ and $\rv$, we get the differential equation $w(\rv)\cos\alpha=w(r)/\sqrt{1+(\partial_x r)^2}$ which can be integrated over the variables $x$ and $r$. So we find 

\begin{equation}\label{l+}
\ell=\frac{1}{\sqrt{\s}}{\cal L}^+(\alpha,v)
\,,
\end{equation}
with $v=\s\rv^2$. The function ${\cal L}^+$ is as defined in Appendix A.

To compute the energy of the string, we first reduce the integral over $x$ in $S$ to that over $r$. This can be done by using the first integral \eqref{In}. Since the resulting integral is divergent at $r=0$, we regularize it by imposing a cutoff $\epsilon$ such that $r\geq\epsilon$. Finally, the regularized expression is given by 

\begin{equation}\label{e+}
E_{R}=\frac{S_R}{T}=\g\sqrt{\frac{\s}{v}}\int^1_{\sqrt{\tfrac{\s}{v}}\epsilon}\,\frac{du}{u^2}\,\ep^{v u^2}\Bigl(1-\cos^2{}\hspace{-1mm}\alpha\,v^4\,\ep^{2v(1-u^2)}\Bigr)^{-\frac{1}{2}}
\,.
\end{equation}
In the limit as $\epsilon\rightarrow 0$, it behaves like 

\begin{equation}\label{E+R}
E_R=\frac{\g}{\epsilon}+E+O(\epsilon)\,.
\end{equation}
Subtracting the $\tfrac{1}{\epsilon}$ term and letting $\epsilon=0$, we get a finite result

\begin{equation}\label{E+}
E=\g\sqrt{\s}\,{\cal E}^+(\alpha,v)+c
\,,
\end{equation}
where $c$ is a normalization constant and ${\cal E}^+$ is the function defined in Appendix A.

At $\alpha=\frac{\pi}{2}$ the string becomes straight. In this case, the expression \eqref{E+} reduces to  

\begin{equation}\label{E90}
E=\g\sqrt{\s}{\cal Q}(v)+c
\,,
\end{equation}
with the function ${\cal Q}$ given by \eqref{Q}. From this it follows that the energy of the string shown in Figure \ref{ng}(b) is simply 

\begin{equation}\label{E|}
E=\g\sqrt{\s}\bigl({\cal Q}(\bar v)-{\cal Q}(v)\bigr)
\,.
\end{equation}
Here $\bar v=\s\rvb^2$. 

\section{Some details on the quark-antiquark potential}
\renewcommand{\theequation}{C.\arabic{equation}}
\setcounter{equation}{0}

In this Appendix we give a brief summary of the basic results about the heavy quark-antiquark potential (the ground state energy of a static quark-antiquark pair) in the presence of two light flavors of equal mass. These are relevant for our discussion in Sec.III. For standard explanations, see \cite{az1,a-strb1} whose conventions we follow, unless otherwise stated. 

In the problem at hand there are two static string configurations: a connected configuration like that at bottom of Figure \ref{c4}(a) and a disconnected configuration similar to that of Figure \ref{c4}(b). In five dimensions the connected configuration includes a string attached to the heavy quark sources on the boundary of five-dimensional space (see Figure \ref{con-ab}(a)). For a Nambu-Goto string in the background geometry \eqref{metric}, the relation between the string energy and quark separation along the $x$-axis is written in parametric form 

\begin{equation}\label{EQQb}
\ell= \frac{2}{\sqrt{\s}}{\cal L}^+(0,v)
 \,,
\quad
E_{\QQb}=2\g\sqrt{\s}\,{\cal E}^+(0,v)+2c\,,
\end{equation}
where $c$ is a normalization constant and $v$ is a dimensionless parameter running from $0$ to $1$. It is given by $v=\s r_0^2$, with $r_0$ a turning point \cite{az1}.

The behavior of $E_{\QQb}$ for small $\ell$ is given by 

\begin{equation}\label{EQQb-small}
E_{\QQb}(\ell)=-\frac{\alpha_{\QQb}}{\ell}+2c+\boldsymbol{\sigma}_{\QQb}\ell +o(\ell)
\,,
\end{equation}
with 
\begin{equation}\label{alpha-QQb}
\alpha_{\QQb}=(2\pi)^3\Gamma^{-4}\bigl(\tfrac{1}{4}\bigr)\g
	\,,\qquad
	\boldsymbol{\sigma}_{\QQb}=\oh(2\pi)^{-2}\Gamma^{4}\bigl(\tfrac{1}{4}\bigr)\g\s
	\,.
\end{equation}
On the other hand, for large $\ell$ it is 

\begin{equation}\label{EQQb-large}
E_{\QQb}(\ell)=\sigma\ell-2\g\sqrt{\s}\,I_0+2c+o(1)
\,,\qquad
\text{with}
\qquad
\sigma=\ep\g\s 
\,.
\end{equation}
Here $\sigma$ is the physical string tension and $I_0$ is defined in Appendix A. Note that the coefficients $\boldsymbol{\sigma}_{\QQb}$ and $\sigma$ are not equal to each other. Numerically, $\boldsymbol{\sigma}_{\QQb}/\sigma\approx 0.805$.

The five-dimensional counterpart of the disconnected configuration is shown in Figure \ref{con-ab}(b). Since the mesons are non-interacting, the energy is twice the heavy-meson mass. The latter is given by Eq.\eqref{Qqb}. 

\begin{figure}[htbp]
\centering
\includegraphics[width=8.55cm]{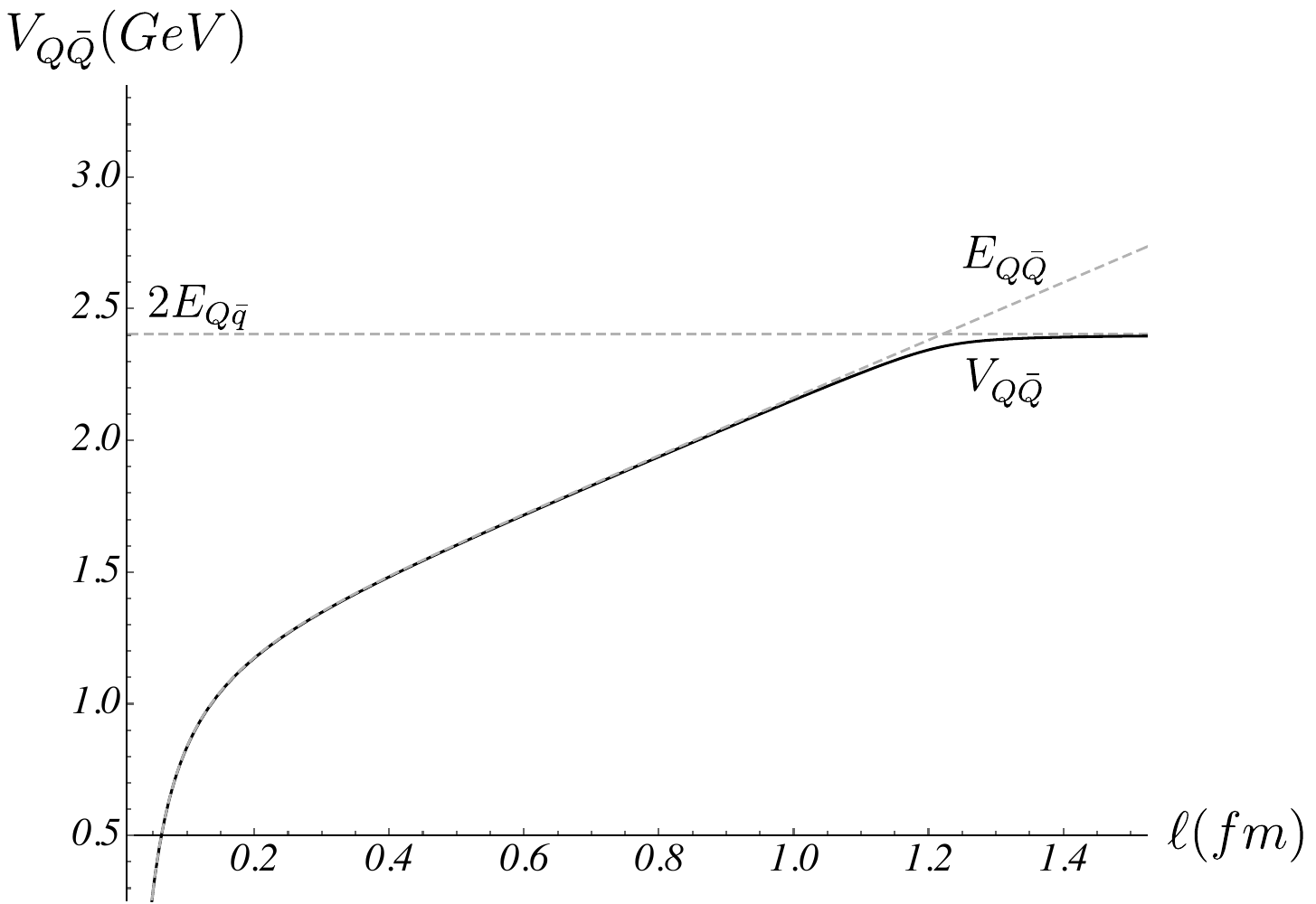}
\caption{{\small The static quark-antiquark potential determined using the model Hamiltonian \eqref{HD-QQb}. Here $\Theta_{\QQb}=47\,\text{MeV}$, as in \cite{bulava}.}}
\label{VQQ}
\end{figure}

As in lattice gauge theory \cite{bulava}, the critical separation (string breaking distance) is defined by equating the energies of the configurations

\begin{equation}\label{LcQQb}
E_{\QQb}(\ell_{\QQb})=2E_{\Qqb}
\,.
\end{equation}
The physical meaning of such a distance is that it gives a condition for determining which configuration is dominant in the ground state of the system (see Figure \ref{VQQ}). The equation simplifies for large quark separations, where $E_{\QQb}(\ell)$ is a linear function of $\ell$.\footnote{For the parameter values we use, this is true for $\ell \gtrsim 0.5\,\text{fm}$, whereas the string breaking distance is about $1\,\text{fm}$.} If so, then it follows from Eqs.\eqref{Qqb} and \eqref{EQQb-large} that the string breaking distance is  

\begin{equation}\label{LQQb-large}
\ell_{\QQb}=\frac{2}{\ep\sqrt{\s}}
\Bigl(
{\cal Q}(q)+\n\frac{\ep^{\oh q}}{\sqrt{q}}+I_0
\Bigr)
\,.
\end{equation}
Here $q$ is a solution to Eq.\eqref{q}. 

The potential is formally defined by requiring $V_{\QQb}=\min\bigl(E_{\QQq},2E_{\Qqb}\bigr)$.\footnote{Conventionally, $V_0$ is used to denote the potential (ground state energy), but we reserve this notation for the $Q\bar Qq\bar q$ system.} Thus it interpolates between $E_{\QQb}$ at small quark separations and $2 E_{\Qqb}$ at large ones. The problem with this formal definition is that it does not say precisely what happens at intermediate quark separations. A way out would be to use the same mixing analysis as in lattice gauge theory \cite{FK,bulava}. So, consider a model Hamiltonian of a two-state system   

\begin{equation}\label{HD-QQb}
{\cal H}(\ell)=
\begin{pmatrix}
E_{\QQb}(\ell) & \Theta_{\QQb} \\
\Theta_{\QQb} & 2E_{\Qqb} \\
\end{pmatrix}
\,,
\end{equation}
with $\Theta_{\QQb}$ describing the mixing between the two states. The potential is given by the smallest eigenvalue of the model Hamiltonian. Explicitly, 

\begin{equation}\label{VQQb}
V_{\QQb}=\oh\Bigl(E_{\QQb}+2E_{\Qqb}\Bigr)
-
\sqrt{\frac{1}{4}\Bigl(E_{\QQb}-2E_{\Qqb}\Bigr)^2+\Theta_{\QQb}^2}
\,.	
\end{equation}

We conclude by giving a simple example of the potential to illustrate this construction. For constant $\Theta_{\QQb}$, the potential is as shown in Figure \ref{VQQ}.
It asymptotically approaches $E_{\QQq}$ as $\ell$ tends to zero and $2E_{\Qqb}$ as $\ell$ tends to infinity. The transition between these two regimes occurs around $\ell=\ell_{\QQb}$, as expected.



\small
\begin{thebibliography}{99}
\bibitem{X38}
Belle Collaboration, S.-K. Choi {\it et al}., Phys. Rev. Lett. {\bf 91}, 262001 (2003). 
\bibitem{lebed}
R. F. Lebed {\it et al}., Summary of Topical Group on Hadron Spectroscopy (RF07) Rare Processes and Precision Frontier of Snowmass 2021, arXiv:2207.14594 [hep-ph].
\bibitem{book}
A. Ali, L. Maiani, and A.D. Polosa, Multiquark Hadrons, Cambridge University Press, 2019.
\bibitem{belle}
Belle Collaboration, A. Bondar {\it et al}., Phys.Rev.Lett. {\bf 108}, 122001 (2012).
\bibitem{bo}
M. Born and J.R. Oppenheimer, Annalen der Physik {\bf 389}, 457 (1927).
\bibitem{braat}
E. Braaten, C. Langmack, and D.H. Smith, Phys.Rev.D {\bf 90}, 014044 (2014). 
\bibitem{QQQQ}
Some of the early works are due to A.M. Green, J. Lukkarinen, P. Pennanen, and C. Michael, Phys.Rev.D {\bf 53}, 261 (1996); F. Okiharu, H. Suganuma, and T.T. Takahashi, Phys.Rev.D {\bf 72}, 014505 (2005).
\bibitem{AP}
P. Bicudo, A. Peters, S. Velten, and M. Wagner, Phys.Rev.D {\bf 103}, 114506 (2021); A. Peters, Investigation of heavy-light four-quark systems by means of Lattice QCD, PhD thesis, Universit\" at Frankfurt am Main (2017).
\bibitem{SP0}
S. Prelovsek, H. Bahtiyar, and J. Petkovic, Phys.Lett.B {\bf 805}, 135467 (2020); M. Sadl and S. Prelovsek, Phys.Rev.D {\bf 104}, 114503 (2021).
\bibitem{uaw}
J. Casalderrey-Solana, H. Liu, D. Mateos, K. Rajagopal, and U.A. Wiedemann, Gauge/String Duality, Hot QCD and Heavy Ion Collisions, Cambridge University Press, 2014.
\bibitem{a-3q0}
O. Andreev, Phys.Rev.D {\bf 78}, 065007 (2008).
\bibitem{coba}
J. Sonnenschein and D. Weissman, Nucl.Phys.B {\bf 920}, 319 (2017). 
\bibitem{a-QQq}
O. Andreev, J. High Energy Phys. {\bf 05} (2021) 173.
\bibitem{QQqq}
O. Andreev, Phys.Rev.D {\bf 105}, 086025 (2022).
\bibitem{FK}
The well-understood example of such an analysis is the heavy quark-antiquark potential in the presence of light dynamical quarks. See, for example, the book by F. Knechtli, M. G\"unther and M. Peardon, Lattice Quantum Chromodynamics, Springer, 2017. 
\bibitem{XA}
X. Artru, Phys.Rept. {\bf 97}, 147 (1983); N. Isgur and J.E. Paton, Phys.Rev.D {\bf 31}, 2910 (1985).
\bibitem{RV}
G.C. Rossi and G. Veneziano, Nucl.Phys.B {\bf 123}, 507 (1977).
\bibitem{VL}
T.T. Takahashi, H. Matsufuru, Y. Nemoto,  and H. Suganuma, Phys.Rev.Lett. {\bf 86}, 18 (2001); Phys.Rev.D {\bf 65}, 114509 (2002); DIK Collaboration, V.G. Bornyakov {\it et al.}, Phys.Rev.D {\bf 70}, 054506 (2004). 
\bibitem{a-strb2}
O. Andreev, Phys.Rev.D {\bf 101}, 106003 (2020).
\bibitem{az1}
O. Andreev and V.I. Zakharov, Phys.Rev.D {\bf 74}, 025023 (2006).
\bibitem{white}
C.D. White, Phys.Lett.B {\bf 652}, 79 (2007).
\bibitem{a-hyb}
O. Andreev, Phys.Rev.D {\bf 86}, 065013 (2012).
\bibitem{a-3qPRD} 
 O. Andreev, Phys.Rev.D {\bf 93}, 105014 (2016).
\bibitem{witten}
E. Witten, J. High Energy Phys. {\bf 9807}, 006 (1998).
\bibitem{son}
J. Erlich, E. Katz, D.T. Son, and M.A. Stephanov, Phys.Rev.Lett. {\bf 95}, 261602 (2005).
\bibitem{voloshin}
M.B. Voloshin, Deciphering the XYZ States, a talk at "{\it 17th Conference on Flavor Physics and CP Violation (FPCP 2019)}, arXiv:1905.13156 [hep-ph].
\bibitem{a-strb1}
O. Andreev, Phys.Lett.B {\bf 804} (2020) 135406.
\bibitem{a-q2}
O. Andreev, Phys.Rev.D {\bf 73}, 107901 (2006).
\bibitem{bulava}
J. Bulava, B. H\"orz, F. Knechtli, V. Koch, G. Moir, C. Morningstar, and M. Peardon, Phys.Lett.B {\bf 793} (2019) 493.
\bibitem{a-3q}
O. Andreev, Phys.Lett.B {\bf 756}, 6 (2016).
\bibitem{kuti}
K.J. Juge, J. Kuti, and C. Morningstar, Phys.Rev.Lett. {\bf 90}, 161601 (2003). 
\bibitem{bulava2}
R. Brett, J. Bulava, D. Darvish, J. Fallica, A. Hanlon, B. H\"orz, and C. Morningstar, AIP Conf. Proc. {\bf 2249}, 030032 (2020).
\bibitem{wagner}
P. Bicudo, M. Cardoso, A. Peters, M. Pflaumer, and M. Wagner, Phys.Rev.D {\bf 96}, 054510 (2017).
 \bibitem{ali}
A. Ali, J.S. Lange, and S. Stone, Prog.Part.Nucl.Phys. {\bf 97}, 123 (2017).
\bibitem{a-stb3q}
O. Andreev, Phys.Rev.D {\bf 104}, 026005 (2021).
\bibitem{DE}
D. Ebert, R.N. Faustov, V.O. Galkin, and W. Lucha, Phys.Rev.D {\bf 76}, 114015 (2007). 
\bibitem{Gi}
M.V. Carlucci, F. Giannuzzi, G. Nardulli, M. Pellicoro, and S. Stramaglia, Eur.Phys.J.C {\bf 57} (2008) 569.
\bibitem{lu}
Q.-F. L\"u and Y.-B. Dong, Phys.Rev.D {\bf 94}, 094041 (2016).
 \bibitem{pionR}
C. Alexandrou {\it et al.} (Extended Twisted Mass Collaboration), The scalar, vector and tensor form factors for the pion and kaon from lattice QCD, arXiv:2111.08135 [hep-lat].
\bibitem{PDG}
P. A. Zyla {\it et al.} (Particle Data Group), PTEP {\bf 2020}, 083C01 (2020).
\bibitem{malda}
J.M. Maldacena, Phys.Rev.Lett. {\bf 80}, 4859 (1998).
\end{thebibliography}
\end{document}